\documentclass[aps,groupedaddress]{revtex4}
\usepackage{epsfig}
\usepackage{graphicx}
\usepackage[T1]{fontenc}
\usepackage{ae}
\usepackage{color}
\usepackage[latin1]{inputenc}
\usepackage{amssymb,amsbsy,amsmath}
\usepackage{bbm}
\usepackage{undertilde}

\newtheorem{lemma}{Lemma}

\newtheorem{prop}{Proposition}
\newtheorem{ass}{Assumption}
\newtheorem{theorem}{Theorem}

\begin{document}

%\begin{center}\today\end{center}

%%%%%%%%%%%%%%%%%%%%%%%%%%%%%%%%%%%%%%%%%%%%%%%%%%%%%%%%%%%%%%%%%%%%%%%%%%%%%

\title[Framework for Jarzynski equation]
{A framework for sequential measurements and general Jarzynski equations}

\author{Heinz-J\"urgen Schmidt and Jochen Gemmer
}
\address{ Universit\"at Osnabr\"uck,
Fachbereich Physik,
 D - 49069 Osnabr\"uck, Germany
}

%\tableofcontents

\begin{abstract}
We formulate a statistical model of two sequential measurements and prove a so-called J-equation that leads to various diversifications
of the well-known Jarzynski equation including the Crooks dissipation theorem.
Moreover, the J-equation entails formulations of the Second Law going back to Wolfgang Pauli.
We illustrate this by an analytically solvable example of sequential discrete position-momentum measurements accompanied with the increase of Shannon entropy.
The standard form of the J-equation extends the domain of applications of the quantum Jarzynski equation in two respects: It includes systems that are initially only in local equilibrium and it extends this equation to the cases where the local equilibrium is described by microcanononical, canonical or grand canonical ensembles. Moreover, the case of a periodically driven quantum system in thermal contact with a heat bath is shown to be covered by the theory presented here.
Finally, we shortly consider the generalized Jarzynski equation in classical statistical mechanics.
\end{abstract}

\maketitle

%%%%%%%%%%%%%%%%%%%%%%%%%%%%%%%%%%%%%%%%%%%%%%%%%%%%%%%%%%%%%%%%%%%%%%%%%%%%%%%%%%%%%%%%%%%%%%%%%%%%%%%%%%%%%%%%%%%%%%%%%%%%%%
\section{Introduction}\label{sec:I}
%%%%%%%%%%%%%%%%%%%%%%%%%%%%%%%%%%%%%%%%%%%%%%%%%%%%%%%%%%%%%%%%%%%%%%%%%%%%%%%%%%%%%%%%%%%%%%%%%%%%%%%%%%%%%%%%%%%%%%%%%%%%%%%

The famous Jarzynski equation represents one of the rare exact results in non-equilibrium statistical mechanics. It is a statement
about the expectation value of the exponential of the work $\left\langle e^{-\beta \,w}\right\rangle $
performed on a system that is initially in thermal equilibrium with
inverse temperature $\beta$, but can be far from equilibrium after the work process. This equation has been first formulated for classical
systems \cite{J97} and subsequently proven to hold for quantum systems \cite{K00}-\cite{M03}. Extensions to systems initially in local thermal equilibrium
\cite{T00}, micro-canonical ensembles \cite{TMYH13} and grand canonical ensembles \cite{SS07} - \cite{YKT12} have been published.
The literature on the Jarzynski equation and its applications is abundant; a concise review is given in \cite{CHT11} with the emphasis
on the connection with other fluctuation theorems.
The most common approach to the quantum Jarzynski equation is in terms of sequential measurements.
This approach will also be adopted in the present paper. The ``work" that appears in the Jarzynski equation is then understood
in terms of the energy differences according to two sequential measurements and hence as a random variable.
Although ``work" is not an observable \cite{TLH07} in the sense of a self-adjoint operator giving rise to a projection-valued measure it can be viewed as a generalized observable \cite{BLPY16} in the sense of a positive-operator-valued measure, see \cite{RCP14} and Subsection \ref{sec:QG}.

Interestingly one can derive from the Jarzynski equation certain inequalities that resemble the $2^{nd}$ law, see, e.~g.~, \cite{CH11}.
However, a closer inspection shows that these inequalities are not exactly statements about the non-decrease of entropy.
Only in the limit case where the system is approximately in thermal equilibrium also {\em after} the work process this
interpretation would be valid. On the other hand there are numerous attempts to derive a  $2^{nd}$ law in the sense of
non-decreasing entropy in quantum mechanics, starting with the paper of W. Pauli \cite{P28} ``on the H-theorem
concerning the increase of entropy in the view of the new quantum mechanics". It is the aim of the present paper
to unify these two routes of research and to identify its common roots.

The structure of the paper is as follows. In Section \ref{sec:SM} we develop a general framework for sequential measurements
and prove a so-called J-equation essentially based on the assumption of a (modified) doubly stochastic conditional probability
matrix. The J-equation depends on an arbitrary sequence $q(j)$ of hypothetic probabilities, but in this paper we will only consider
two special cases, the case R (``real probabilities") and the case S (``standard probabilities") to be defined in Section \ref{sec:Q}.
In the case R the J-equation implies, via Jensen's inequality,
an increase of the (modified) Shannon entropy from the first to the second measurement.
In the next Section  \ref{sec:Q} we specialize the general framework of Section \ref{sec:SM} to the case of quantum theory
such that the first measurement is of L\"uders type satisfying two more assumptions. Then we can reformulate the J-equation
for the case S where the initial density matrix is a function ${\mathcal G}$ of $L$ commuting self-adjoint operators, see Theorem \ref{T1}.
Special choices for the function ${\mathcal G}$ and the $L$ commuting self-adjoint operators lead to various diversifications of the
Jarzynski equation: the local equilibrium given by $N$ canonical ensembles (subsection \ref{sec:L}),
the micro-canonical ensemble (subsection \ref{sec:M}) and the  grand canonical ensemble case (subsection \ref{sec:R}).
Moreover, in subsection \ref{sec:PT} we consider recently discovered cases of a periodically driven quantum system that
are in quasi-equilibrium with a heat bath possessing a quasi-temperature $1/\vartheta$
and show how these cases can also be covered by the present theory.
In all these applications we will obtain case S variants of the $2^{nd}$ law-like statements following from the Jarzynski equations
via Jensen's inequality.

The above-mentioned case R variant of the $2^{nd}$ law also holds in quantum theory.
This will be discussed in some more detail in Section \ref{sec:S} containing further applications.
It will be instructive to consider the analytically solvable example
of two subsequent discrete position-momentum measurements at a free particle moving in one dimension
and to confirm the mentioned increase of Shannon entropy, see subsection \ref{sec:SA}.
In the following subsection \ref{sec:F} we show how to integrate the quantum version of the Crooks dissipation theorem into our approach.
We briefly discuss how the results hitherto derived can be transferred to the classical realm in Section \ref{sec:C}.
We close with a summary and outlook in Section \ref{sec:SO}.

%%%%%%%%%%%%%%%%%%%%%%%%%%%%%%%%%%%%%%%%%%%%%%%%%%%%%%%%%%%%%%%%%%%%%%%%%%%%%%%%%%%%%%%%%%%%%%%%%%%%%%%%%%%%%%%%%%%%%%%%%%%%%%
\section{Statistical model of sequential measurements}\label{sec:SM}
%%%%%%%%%%%%%%%%%%%%%%%%%%%%%%%%%%%%%%%%%%%%%%%%%%%%%%%%%%%%%%%%%%%%%%%%%%%%%%%%%%%%%%%%%%%%%%%%%%%%%%%%%%%%%%%%%%%%%%%%%%%%%%%

%%%%%%%%%%%%%%%%%%%%%%%%%%%%%%%%%%%%%%%%%%%%%%%%%%%%%%%%%%%%%%%%%%%%%%%%%%%%%%%%%%%%%%%%%%%%%%%%%%%%%%%%%%%%%%%%%%%%%%%%%%%%%%
\subsection{Simple case}\label{sec:SMS}
%%%%%%%%%%%%%%%%%%%%%%%%%%%%%%%%%%%%%%%%%%%%%%%%%%%%%%%%%%%%%%%%%%%%%%%%%%%%%%%%%%%%%%%%%%%%%%%%%%%%%%%%%%%%%%%%%%%%%%%%%%%%%%%

We consider two sequential measurements at the same physical system at times $t_0<t_1$ with respective outcome sets ${\mathcal I}$ and ${\mathcal J}$.
These sets are assumed to be finite or countably infinite. Hence the joint outcome of the two measurements can be represented
by the pair $(i,j)\in {\mathcal I}\times {\mathcal J}$. We define
\begin{equation}\label{SM1}
 {\sf E}\equiv {\mathcal I}\times {\mathcal J}
\end{equation}
as the set of ``elementary events" and describe the probability of elementary events by a function
\begin{equation}\label{SM2}
P:{\sf E}\rightarrow [0,1]
\end{equation}
subject to the natural condition
\begin{equation}\label{SM3}
  \sum_{(i,j)\in{\sf E}}P(i,j)=1
  \;.
\end{equation}
As usual, one defines the first and second marginal probability functions
\begin{eqnarray}\label{SM4a}
  p&:&{\mathcal I}\rightarrow [0,1]\\
  \label{SM4b}
  p(i)&\equiv& \sum_{j\in{\mathcal J}}P(i,j)
  \;,
\end{eqnarray}
and
\begin{eqnarray}\label{SM5a}
  \hat{p}&:&{\mathcal J}\rightarrow [0,1]\\
  \label{SM5b}
 \hat{p}(j)&\equiv& \sum_{i\in{\mathcal I}}P(i,j)
  \;.
\end{eqnarray}
For sake of simplicity we will assume
\begin{equation}\label{SM6}
  p(i)>0 \mbox{ for all } i\in{\mathcal I}
  \;.
\end{equation}
This could be achieved by deleting all outcomes $i\in{\mathcal I}$ with $p(i)=0$ thereby reducing the set ${\mathcal I}$.
Due to (\ref{SM6}) the ``conditional probability"
\begin{equation}\label{SM7}
 \pi(j|i)\equiv \frac{P(i,j)}{p(i)}
\end{equation}
can be defined for all $(i,j)\in{\sf E}$. It satisfies
\begin{equation}\label{SM8}
  \sum_{j\in{\mathcal J}}\pi(j|i)=1 \mbox{ for all }i\in{\mathcal I}
  \;,
\end{equation}
and hence can be considered as a stochastic matrix. Note further that
\begin{equation}\label{SM8a}
  \hat{p}(j)=\sum_{i\in{\mathcal I}}\pi(j|i)\,p(i)
  \;.
\end{equation}
If additionally $\pi$ is a ``doubly stochastic matrix", i.~e.~,
\begin{equation}\label{SM9}
  \sum_{i\in{\mathcal J}}\pi(j|i)=1 \mbox{ for all }j\in{\mathcal J}
  \;,
\end{equation}
the  triple $({\mathcal I}, {\mathcal J}, P)$ will be called a ``statistical model of two sequential measurements" ($SM^2$).

In accordance with the usual nomenclature of probability theory, functions $X:{\sf E}\rightarrow {\mathbbm R}$ are also called ``random variables".
Their expectation value is defined as
\begin{equation}\label{SM9a}
 \langle X\rangle \equiv \sum_{(i,j)\in {\sf E}} X(i,j)\,P(i,j)
 \;,
\end{equation}
if the series converges.
Using a sloppy notation the expectation value will be sometimes also written as $\langle X(i,j)\rangle$ if no misunderstanding is likely to occur.
We have the following result:
\begin{prop}\label{Prop1}
If $({\mathcal I}, {\mathcal J}, P)$  is an $SM^2$ and $q:{\mathcal J}\rightarrow [0,1]$ a sequence satisfying
\begin{equation}\label{P1a}
  \sum_{j\in{\mathcal J}}q(j)=1
  \;,
\end{equation}
then
\begin{equation}\label{P1b}
 \left\langle \frac{q(j)}{p(i)}\right\rangle=1
 \;.
\end{equation}

Conversely, if $({\mathcal I}, {\mathcal J}, P)$ satisfies the above conditions, but not necessarily (\ref{SM9}),
and (\ref{P1b}) holds for all  $q:{\mathcal J}\rightarrow [0,1]$ satisfying (\ref{P1a}) then $\pi(j|i)$ will be
doubly stochastic.
\end{prop}

The $q(j)$ will also be called ``hypothetical probabilities" in contrast to the ``real probabilities" $\hat{p}(j)$.
The choice $q(j)=\hat{p}(j)$ will be referred to as the ``case R".
The above proposition essentially says that the J-equation is equivalent to $\pi(j|i)$ being doubly stochastic.
\\

\noindent {\bf Proof} of Proposition \ref{Prop1}:
\begin{eqnarray}
\label{SM10a}
   \left\langle \frac{q(j)}{p(i)}\right\rangle &\stackrel{(\ref{SM9a})}{=}&  \sum_{i,j}\frac{q(j)}{p(i)}\,P(i,j)\\
   \label{SM10b}
   &\stackrel{(\ref{SM7})}{=}& \sum_{j} q(j)\,\sum_{i}\pi(j|i)\\
   \label{SM10c}
   &\stackrel{(\ref{SM9})}{=}&\sum_{j}q(j)\stackrel{(\ref{P1a})}{=}1
   \;.
\end{eqnarray}
For the converse statement choose $j'\in{\mathcal J}$ arbitrarily and let $q(j)=\delta_{j,j'}$. Then
\begin{equation}\label{SM10d}
 1=\left\langle \frac{q(j)}{p(i)}\right\rangle=\sum_{i,j}q(j)\, \pi(j|i) =\sum_{i,j}\delta_{j,j'}\, \pi(j|i) =
 \sum_i \pi(j'|i)
 \;,
\end{equation}
which means that $\pi(j|i)$ will be doubly stochastic.
\hfill$\Box$\\

We will call Eq.~(\ref{P1b}) and its modified form (\ref{P2b}) the ``J-equation" since we think that it contains
the probabilistic core of the Jarzynski equation but should be distinguished from the latter for the sake of clarity.
To illustrate this claim we note that any sequence
$p:{\mathcal I}\rightarrow [0,1]$ of probabilities satisfying (\ref{SM6}) may be written in the form
\begin{equation}\label{SM11}
 p(i)= \exp\left( -\beta\left( E_i-F\right)\right)
 \;,
\end{equation}
and, analogously,
\begin{equation}\label{SM12}
q(j)= \exp\left( -\beta\left( E'_j-F'\right)\right)
 \;,
\end{equation}
where the $\beta, E_i, E'_j, F,F'$ are certain real parameters, not uniquely determined by (\ref{SM11}) and (\ref{SM12}).
Then  Eq.~(\ref{P1b}) can be written as
\begin{equation}\label{SM13}
  \left\langle e^{-\beta\,w}\right\rangle=e^{-\beta \Delta F}
  \;,
\end{equation}
where $w:{\sf E}\rightarrow {\mathbbm R}$ is a random variable defined by $w(i,j)\equiv E'_j- E_i$ and
$\Delta F\equiv F'-F$. Indeed, Eq.~(\ref{SM13}) has the form of the standard Jarzynski equation, but, in general,
the parameters occurring in (\ref{SM13}) will not have the physical meaning of inverse temperature $\beta$,
work $w$, and difference of free energies $\Delta F$, as required for the Jarzynski equation.
Even in the special case where the usual physical interpretation of the parameters $\beta, E_i, E'_j, F,F'$ holds,
we have not yet proven the standard Jarzynski equation, because we still would have to confirm the conditions
of Proposition \ref{Prop1} for this special case.\\

Let  $({\mathcal I}, {\mathcal J}, P)$  be an $SM^2$  and choose $q(j)=\hat{p}(j)$ for all $j\in{\mathcal J}$, that is,
replace the hypothetical probabilities by the real ones. Then, by Proposition \ref{Prop1},
\begin{equation}\label{SM14}
 \left\langle \frac{\hat{p}(j)}{p(i)}\right\rangle=1
 \;.
\end{equation}
Since the logarithm (with arbitrary basis) is a concave function, Jensen's inequality yields
\begin{equation}\label{SM15}
  \langle \log\,X\rangle \le \log\,\langle X\rangle
\end{equation}
for any random variable $X:{\sf E}\rightarrow{\mathbbm R}$. It follows that
\begin{eqnarray}
\label{SM16a}
  0 &=& \log\,1\stackrel{(\ref{SM14})}{=}\log\,  \left\langle \frac{\hat{p}(j)}{p(i)}\right\rangle\\
  \label{SM16b}
   &\stackrel{(\ref{SM15})}{\ge}& \left\langle\log\, \frac{\hat{p}(j)}{p(i)}\right\rangle \\
   \label{SM16c}
   &=& \left\langle\log\, \hat{p}(j)\right\rangle-\left\langle\log\, p(i)\right\rangle\\
   \label{SM16d}
   &=& \sum_{j\in{\mathcal J}} \hat{p}(j)\,\log  \hat{p}(j) \,-\,  \sum_{i\in{\mathcal I}} p(i)\,\log  p(i)
   \;.
\end{eqnarray}
In other words, the Shannon entropy
\begin{equation}\label{SM17}
 S(p)=-\sum_i p_i\,\log p_i
 \;,
\end{equation}
usually defined for the logarithm with basis $2$, see \cite{S48},
does not decrease in the statistical model of two sequential measurements, i.~e.~,
\begin{equation}\label{SM18}
  S(\hat{p})\ge S(p)
  \;.
\end{equation}

It is an obvious question under which circumstances the inequality in ({\ref{SM18}) will be a strict one. We will answer this question
only for the case of finite ${\mathcal I}={\mathcal J}$:
\begin{prop} \label{Prop2}
  Let ${\mathcal I}={\mathcal J}$ and $\left| {\mathcal I }\right|=n$. Then $S(\hat{p})=S(p)$ iff the conditional probability is of
  permutational type, i.~e., iff $\pi(j|i)=\delta_{j,\sigma(i)}$ for some permutation $\sigma\in {\mathcal S}(n)$.
\end{prop}
\noindent {\bf Proof}: The if-part follows since the sum (\ref{SM17}) is invariant under permutations. For the only-if part
we invoke the theorem of Birkhoff-von Neumann \cite{B46} saying that any doubly stochastic matrix is the convex sum of permutational
matrices. Assume that $\pi$ is not of permutational type and hence the convex sum will be a proper one. This means that $\pi$ can be written in the form
\begin{equation}\label{Pa}
 \pi=\sum_{\mu=1}^{M}\lambda_\mu\,\hat{\sigma}_\mu\;,
\end{equation}
such that $\lambda_\mu > 0$ for all $\mu=1,\ldots,M$, $M>1$, $\sum_{\mu}\lambda_\mu=1$, and
\begin{equation}\label{Pb}
  \hat{\sigma}_\mu(j|i)=\delta_{j,\sigma_\mu(i)} \quad\mbox{for some }\sigma_\mu\in  {\mathcal S}(n)
  \;.
\end{equation}
Recall that the function $f(x)\equiv -x\,\log x$ is strictly concave for $x\in [0,1]$. Then we obtain
\begin{eqnarray}
\label{Pc}
  S(\hat{p}) &=& \sum_j f(\hat{p}(j))\\
  \label{Pd}
   &\stackrel{(\ref{SM8a},\ref{Pa})}{=}&  \sum_j f\left(\sum_\mu \lambda_\mu \sum_i \hat{\sigma}_\mu (j|i)p(i) \right)\\
   \label{Pe}
&>& \sum_j\sum_\mu \lambda_\mu\, f\left(\sum_i \hat{\sigma}_\mu (j|i)p(i)  \right)\\
\label{Pf}
&\stackrel{(\ref{Pb})}{=}& \sum_j\sum_\mu \lambda_\mu\, f\left(\sum_i \delta_{j,\sigma_\mu(i)} p(i)  \right)\\
\label{Pg}
&=& \sum_j\sum_\mu \lambda_\mu\, f\left(p\left( \sigma_\mu^{-1}(j)\right) \right)\\
\label{Ph}
&=& \left(\sum_\mu \lambda_\mu\right) \left( \sum_i f(p(i))\right)\\
\label{Pi}
&=& S(p)
\;,
\end{eqnarray}
where in (\ref{Pe}) we have applied Jensen's inequality using that $f$ is strictly concave and the convex sum (\ref{Pa}) is a proper one.
Summarizing, $S(\hat{p})> S(p)$ if $\pi$ is not of permutational type, which completes the proof of Prop. \ref{Prop2}.
\hfill$\Box$\\

One may ask which assumption is responsible for the asymmetry between the two sequential measurements that appears in (\ref{SM18}).
Obviously this is the property (\ref{SM9}) of the conditional probability matrix being doubly stochastic that is
only postulated for the first conditional probability $\pi$ and not for the second one $\hat{\pi}(i|j)\equiv\frac{P(i,j)}{\hat{p}(j)}$.
It will be instructive to consider the situation in which both matrices, $\pi$ and $\hat{\pi}$, are doubly stochastic.
For sake of simplicity we will assume that the outcome sets ${\mathcal I}$ and ${\mathcal J}$ are finite,  both containing exactly $n$ elements,
and that $P(i,j)>0$ for all $i\in{\mathcal I}$ and $j\in{\mathcal J}$. It follows that, in matrix notation,
$\pi\,p=\hat{p}$ and $\hat{\pi}\,\hat{p}=p$, cf.~(\ref{SM8a}), moreover the double stochasticity may be written as $\pi\,{\bf 1}=\hat{\pi}\,{\bf 1}={\bf 1}$,
where ${\bf 1}$ denotes the constant vector ${\bf 1}=(1,1,\ldots,1)$. Hence the matrix $\Pi\equiv \hat{\pi}\,\pi$ has the
two positive invariant distributions $\frac{1}{n}{\bf 1}$ and $p$. Since  $P(i,j)>0$ the matrix $\Pi$ is irreducible and hence
its positive invariant distribution is unique (Theorem $54$ of \cite{S12}). Consequently, $p(i)=\hat{p}(j)=\frac{1}{n}$ for all $i\in{\mathcal I}$
and $j\in{\mathcal J}$. This characterizes the constant distribution with maximal Shannon entropy and hence a completely symmetric situation.

%%%%%%%%%%%%%%%%%%%%%%%%%%%%%%%%%%%%%%%%%%%%%%%%%%%%%%%%%%%%%%%%%%%%%%%%%%%%%%%%%%%%%%%%%%%%%%%%%%%%%%%%%%%%%%%%%%%%%%%%%%%%%%
\subsection{Modified case}\label{sec:SMM}
%%%%%%%%%%%%%%%%%%%%%%%%%%%%%%%%%%%%%%%%%%%%%%%%%%%%%%%%%%%%%%%%%%%%%%%%%%%%%%%%%%%%%%%%%%%%%%%%%%%%%%%%%%%%%%%%%%%%%%%%%%%%%%%

Now we will formulate a slightly more general framework for $SM^2$ that is motivated by applications using quantum theory
in Section \ref{sec:Q} and partially follows the account of Wolfgang Pauli in \cite{P28}, Ch. I \S 2.
We assume that the outcome sets ${\mathcal I}$ and ${\mathcal J}$ are divided into disjoint cells
(``Elementarbereiche" in \cite{P28}) such that the probability $P(i,j)$ is constant over the cells.
The construction is similar to the operation of ``coarse graining" in physical theories, but in those cases the probability will
typically not be constant over the cells and the following considerations will be at most approximately valid.
The mentioned cells will be written as the inverse images of suitable maps
\begin{equation}\label{SM19}
 \Pi_{\mathcal I}:{\mathcal I}\rightarrow{\mathcal I}'
 \mbox{ and }
 \Pi_{\mathcal J}:{\mathcal J}\rightarrow{\mathcal J}'
 \;,
\end{equation}
and are assumed to be finite.  ${\mathcal I}'$ and ${\mathcal J}'$ can hence be viewed as the respective sets of cells.
We define the cell sizes
\begin{eqnarray}
\label{SM20a}
  d: {\mathcal I}'\rightarrow{\mathbbm N}, &\quad& d(i')\equiv \left| \Pi_{\mathcal I}^{-1}(i')\right|\;, \\
  \label{SM20b}
  D: {\mathcal J}'\rightarrow{\mathbbm N}, &\quad& D(j')\equiv \left| \Pi_{\mathcal J}^{-1}(j')\right|\;.
\end{eqnarray}
As mentioned above, the probability is assumed to be constant over cells and hence gives rise to a modified probability function
$P':{\mathcal I}'\times {\mathcal J}'\rightarrow[0,1]$ via
\begin{equation}\label{SM21}
 P'(i',j')\equiv d(i')\,D(j')\,P(i,j), \mbox{ if } \Pi_{\mathcal I}(i)=i' \mbox{ and }\Pi_{\mathcal J}(j)=j'
 \;.
\end{equation}
We note that
\begin{equation}\label{SM22}
 1\stackrel{(\ref{SM3})}{=}\sum_{i\in{\mathcal I},j\in{\mathcal J}}P(i,j)
 =\sum_{i'\in{\mathcal I}',j'\in{\mathcal J}'}\sum_{i\in\Pi_{\mathcal I}^{-1}(i')}\sum_{j\in\Pi_{\mathcal J}^{-1}(j')}P(i,j)
 =\sum_{i'\in{\mathcal I}',j'\in{\mathcal J}'}d(i')\,D(j')P(i,j)\stackrel{(\ref{SM21})}{=}\sum_{i'\in{\mathcal I}',j'\in{\mathcal J}'}P'(i',j')
 \;,
\end{equation}
as it must hold for a probability function.
Here and in what follows the index $i$ within a sum over $i'$ denotes an arbitrary element of the cell $\Pi_{\mathcal I}^{-1}(i')$, analogous for $j$.
As in subsection \ref{sec:SMS} we define the modified marginal probability and obtain
\begin{equation}\label{SM23}
 p'(i')\equiv \sum_{j'}P(i',j')\stackrel{(\ref{SM21})}{=}\sum_{j'}P(i,j)\,d(i')\,D(j')=\sum_j P(i,j)\,d(i')\stackrel{(\ref{SM4b})}{=}p(i)\,d(i')
 \;.
\end{equation}
Analogously for the modified conditional probability:
\begin{equation}\label{SM24}
 \pi'(j'|i')\equiv \frac{P'(i',j')}{p'(i')}\stackrel{(\ref{SM21},\ref{SM23})}{=}
 \frac{P(i,j) d(i')D(j')}{p(i) d(i')}\stackrel{(\ref{SM7})}{=}\pi(j|i)\,D(j')
 \;.
\end{equation}
The condition of $\pi$ being doubly stochastic entails the following property of $\pi'$:
\begin{equation}\label{SM25}
\sum_{i'}\pi'(j'|i')\,d(i')\stackrel{(\ref{SM24})}{=}\sum_{i'}\pi(j|i)\,D(j')\,d(i')=\sum_i \pi(j|i)\,D(j')\stackrel{(\ref{SM9})}{=}D(j')
\;.
\end{equation}
Next we express the Shannon entropy in terms of the modified probabilities:
\begin{equation}\label{SM26}
 S(p)\stackrel{(\ref{SM17})}{=}-\sum_i p(i)\,\log p(i)\stackrel{(\ref{SM23})}{=}-\sum_i\frac{p'(i')}{d(i')}\log \frac{p'(i')}{d(i')}
 = -\sum_{i'}p'(i') \log  \frac{p'(i')}{d(i')}\equiv S'(p')
 \;,
\end{equation}
cp.~Eq.~(17) of \cite{P28}.\\

When defining the modified framework for sequential measurements we will omit the primes of the preceding equations and
postulate two non-vanishing functions
\begin{equation}\label{SM27}
 d:{\mathcal I}\rightarrow{\mathbbm N} \mbox{ and } D:{\mathcal J}\rightarrow{\mathbbm N}
 \end{equation}
that play the role of the cell sizes considered above. In the next section \ref{sec:Q} the $d(i)$ and $D(j)$ will appear as the degeneracies
of certain eigenspaces.

Then the $5$-tuple $({\mathcal I}, {\mathcal J}, P, d, D)$ will be called a ``modified statistical model
of two sequential measurements" ($mSM^2$), iff the condition of $\pi$ being doubly stochastic is replaced by the unprimed version of (\ref{SM25}):
\begin{equation}\label{SM28}
  \sum_{i\in{\mathcal I}} \pi(j|i)\,d(i)=D(j) \mbox{ for all } j\in{\mathcal J}
  \;.
\end{equation}
We will also denote a conditional probability function $\pi:{\sf E}\rightarrow [0,1]$ satisfying  (\ref{SM28})
as being of ``modified doubly stochastic" type.

Consequently we obtain the following variant of Proposition \ref{Prop1}:
\begin{prop}\label{Prop3}
If $({\mathcal I}, {\mathcal J}, P,d,D)$  is an $mSM^2$ and $q:{\mathcal J}\rightarrow [0,1]$ a sequence satisfying
\begin{equation}\label{P2a}
  \sum_{j\in{\mathcal J}}q(j)=1
  \;,
\end{equation}
then
\begin{equation}\label{P2b}
 \left\langle \frac{d(i)}{D(j)}\frac{q(j)}{p(i)}\right\rangle=1
 \;.
\end{equation}
Conversely, if $({\mathcal I}, {\mathcal J}, P,d,D)$ satisfies the above conditions, but not necessarily (\ref{SM28}),
and (\ref{P2b}) holds for all  $q:{\mathcal J}\rightarrow [0,1]$ satisfying (\ref{P2a}) then $\pi(j|i)$ will be
of modified doubly stochastic type.
\end{prop}
The proof is completely analogous to that of Proposition \ref{Prop1}. Moreover, it follows that the modified Shannon entropy
\begin{equation}\label{SM29}
  S'(p)\equiv -\sum_i p(i)\log \frac{p(i)}{d(i)}
\end{equation}
does not decrease in the modified statistical model of two sequential measurements, i.~e.~,
\begin{equation}\label{SM30}
  S'(\hat{p})\ge S'(p)
  \;,
\end{equation}
where
\begin{equation}\label{SM29a}
  S'(\hat{p})\equiv -\sum_j \hat{p}(j)\log \frac{\hat{p}(j)}{D(j)}
  \;.
\end{equation}

This equation is analogous to the statement $\frac{dS}{dt}\ge 0$ after Eq.~(22) in \cite{P28}
that has been proven by Pauli using the stronger symmetry condition (in our notation)
\begin{equation}\label{SM31}
\frac{\pi(j|i)}{D(j)}=\frac{\pi(i|j)}{d(i)}
\;,
\end{equation}
see Eq.(21) in \cite{P28}, justified by $1^{st}$ order perturbation theory (``Fermi's Golden Rule"). We note that in general
(\ref{SM31}) need not hold, see Section \ref{sec:SA} for a counter-example, but there are also positive examples beyond the Golden Rule,
see subsection \ref{sec:F}.

%%%%%%%%%%%%%%%%%%%%%%%%%%%%%%%%%%%%%%%%%%%%%%%%%%%%%%%%%%%%%%%%%%%%%%%%%%%%%%%%%%%%%%%%%%%%%%%%%%%%%%%%%%%%%%%%%%%%%%%%%%%%%%%
\section{Applications to quantum theory}\label{sec:Q}
%%%%%%%%%%%%%%%%%%%%%%%%%%%%%%%%%%%%%%%%%%%%%%%%%%%%%%%%%%%%%%%%%%%%%%%%%%%%%%%%%%%%%%%%%%%%%%%%%%%%%%%%%%%%%%%%%%%%%%%%%%%%%%%
%%%%%%%%%%%%%%%%%%%%%%%%%%%%%%%%%%%%%%%%%%%%%%%%%%%%%%%%%%%%%%%%%%%%%%%%%%%%%%%%%%%%%%%%%%%%%%%%%%%%%%%%%%%%%%%%%%%%%%%%%%%%%%%
\subsection{General case}\label{sec:QG}
%%%%%%%%%%%%%%%%%%%%%%%%%%%%%%%%%%%%%%%%%%%%%%%%%%%%%%%%%%%%%%%%%%%%%%%%%%%%%%%%%%%%%%%%%%%%%%%%%%%%%%%%%%%%%%%%%%%%%%%%%%%%%%%

We will investigate how the (modified) statistical model of sequential measurements outlined in the preceding subsections
can be realized within the framework of quantum theory. The identification of the respective concepts will
be facilitated by denoting them with the same letters. Additionally to a number of usual assumptions we will
use Assumption \ref{AS1} and \ref{AS2} that are highlighted below.

We consider a quantum system with a Hilbert space ${\mathcal H}$ and a finite number of mutually commuting self-adjoint
operators $\utilde{E}_1,\ldots,\utilde{E}_L$ defined on (suitable domains of) ${\mathcal H}$. They are assumed to have a pure point
spectrum and hence a family of common eigenprojections $(\utilde{P}_i)_{i\in{\mathcal I}}$  such that
\begin{equation}\label{G1}
 \utilde{E}_\lambda=\sum_{i\in{\mathcal I}}E_i^{(\lambda)}\,\utilde{P}_i,\quad \lambda=1,\ldots,L
 \;.
\end{equation}
Here ${\mathcal I}$ is a finite or countable infinite index set to be identified with the outcome set of the first measurement
according to Section \ref{sec:SM}. The $\utilde{P}_i$ are assumed to be of finite degeneracy,
\begin{equation}\label{G2}
  d(i)\equiv \mbox{Tr} \left( \utilde{P}_i\right)<\infty,\quad\mbox{for all }i\in{\mathcal I}
  \;,
\end{equation}
and are chosen as maximal projections in the sense that $i\neq j$ implies $E_i^{(\lambda)}\neq E_j^{(\lambda)}$ for at least one $\lambda=1,\ldots,L$.
Note the completeness relation
\begin{equation}\label{G3}
 \sum_{i\in{\mathcal I}}\utilde{P}_i={\mathbbm 1}
 \;.
\end{equation}
Physically, the $\utilde{E}_1,\ldots,\utilde{E}_L$ correspond to observables that can be jointly measured. We assume a (mixed) state
of the system before the time $t=t_0$ described by a density operator $\rho$ and perform a joint L\"uders measurement, cf.~\cite{BLPY16} (10.22),
of $\utilde{E}_1,\ldots,\utilde{E}_L$
at the time $t=t_0$.
The probability of the outcome $i\in{\mathcal I}$ will be
\begin{equation}\label{G4}
  p(i)=\mbox{Tr} \left( \rho\,\utilde{P}_i\right)
  \;,
\end{equation}
satisfying
\begin{equation}\label{G4a}
 \sum_{i\in{\mathcal I}} p(i)=1
  \;.
\end{equation}
In accordance with (\ref{SM6}) we will make the following
\begin{ass}\label{AS1}
  \begin{equation}\label{AS1a}
    p(i)>0 \mbox{ for all } i\in{\mathcal I}
    \;.
  \end{equation}
\end{ass}
The validity of this assumption could be achieved by restricting the Hilbert space ${\mathcal H}$ to the subspace spanned by the eigenspaces
of those $\utilde{P}_i$ with $p(i)=\mbox{Tr}\left(\rho\,\utilde{P}_i\right)>0$.

After the first measurement of the $\utilde{E}_1,\ldots,\utilde{E}_L$ the system is subject to a further time evolution and a
second measurement of (possibly) other observables. Thus the primary preparation together with the first measurement may be considered
as another preparation of a certain state, in general different from the initial state $\rho$. If a selection according
to a particular outcome $i\in{\mathcal I}$ is involved this state will be, according to the assumption of a L\"uders measurement,  cf.~\cite{BLPY16} (10.22),
\begin{equation}\label{G5a}
  \rho_i=\frac{\utilde{P}_i\,\rho\,\utilde{P}_i}{\mbox{Tr}\left(\rho\,\utilde{P}_i \right)}=\frac{\utilde{P}_i\,\rho\,\utilde{P}_i}{p(i)}
  \;.
\end{equation}
If no selection according to a particular outcome is involved the state resulting after the first measurement will rather be the
mixed state
\begin{equation}\label{G5b}
  \rho_1=\sum_{i\in{\mathcal I}}p(i)\,\rho_i \stackrel{(\ref{G5a})}{=}\sum_{i\in{\mathcal I}}\utilde{P}_i\,\rho\,\utilde{P}_i
  \;.
\end{equation}

In order to apply the results of the preceding section we will make the following crucial assumption
\begin{ass}\label{AS2}
  \begin{equation}\label{AS2a}
    \rho_i=\frac{1}{d(i)}\utilde{P}_i \mbox{ for all } i\in{\mathcal I}
    \;.
  \end{equation}
\end{ass}
If $\utilde{P}_i$ is a one-dimensional projection, i.~e., if $d(i)=1$, the assumption (\ref{AS2a}) will be automatically
satisfied. In the case of $d(i)>1$ this assumption means that $\rho$ is diagonal w.~r.~t.~any common eigenbasis of the
$\utilde{E}_1,\ldots,\utilde{E}_L$. An important case where (\ref{AS2a}) holds is given if $\rho$ is a function of the operators
$\utilde{E}_1,\ldots,\utilde{E}_L$, say,
\begin{equation}\label{G5c}
  \rho={\mathcal G}\left( \utilde{E}_1,\ldots,\utilde{E}_L\right)
\;.
\end{equation}
This has to be interpreted in the sense of functional calculus as
\begin{equation}\label{G5d}
  \rho=\sum_{i\in{\mathcal I}}{\mathcal G}\left(E_i^{\boldsymbol{\lambda}}\right)\,\utilde{P}_i
 \;,
\end{equation}
where $E_i^{\boldsymbol{\lambda}}$ will be the short-hand notation for $\left( E_i^{(1)},\ldots,E_i^{(L)}\right)$.
It follows that, in accordance with (\ref{AS2a}),
\begin{equation}\label{G5e}
  \rho_i\stackrel{(\ref{G5a})}{=}\frac{1}{p(i)}\utilde{P}_i\,\rho\,\utilde{P}_i\stackrel{(\ref{G5d})}{=}
  \frac{1}{p(i)}{\mathcal G}\left(E_i^{\boldsymbol{\lambda}}\right)\utilde{P}_i
  =\frac{1}{d(i)}\utilde{P}_i
  \;,
\end{equation}
since
\begin{equation}\label{G5f}
  p(i)\stackrel{(\ref{G4})}{=}\mbox{Tr}\left( \rho\,\utilde{P}_i\right)
  \stackrel{(\ref{G5d})}{=}{\mathcal G}\left(E_i^{\boldsymbol{\lambda}}\right)\mbox{Tr }\utilde{P}_i
  \stackrel{(\ref{G2})}{=}
  {\mathcal G}\left(E_i^{\boldsymbol{\lambda}}\right)d(i)
  \;.
\end{equation}

In what follows we will refer to the case (\ref{G5c}) as the ``standard case" (case S).
However, we stress that this is not the most general case compatible with the Assumption \ref{AS2}
as the counter-example of all $d(i)=1$ and $\rho$ not commuting with the $\utilde{P}_i$ shows.\\

Next we consider a second set of observables described by the mutually commuting self-adjoint operators $\utilde{F}_1,\ldots,\utilde{F}_L$
subject to analogous assumptions. Hence the following holds:
\begin{equation}\label{G6}
 \utilde{F}_\lambda=\sum_{j\in{\mathcal J}}F_j^{(\lambda)}\,\utilde{Q}_j,\quad \lambda=1,\ldots,L
 \;,
\end{equation}
\begin{equation}\label{G7}
  D(j)\equiv \mbox{Tr} \left(\utilde{Q}_j\right)<\infty,\quad\mbox{for all }j\in{\mathcal J}
  \;,
\end{equation}
and
\begin{equation}\label{G8}
 \sum_{j\in{\mathcal J}}\utilde{Q}_j={\mathbbm 1}
 \;.
\end{equation}
We have chosen another index set ${\mathcal J}$ for the second set of observables in order to stress that no natural identification between both index sets
is required in what follows. Obviously,  ${\mathcal J}$ has to be identified with the second outcome set introduced in Section \ref{sec:SM}.
In general the $\utilde{E}_\lambda$ will not commute with the $\utilde{F}_\mu$. We assume that
a second measurement of the  $\utilde{F}_1,\ldots,\utilde{F}_L$ will be performed at the time $t=t_1>t_0$, not necessarily of L\"uders type.
Between the two measurements in the time interval $(t_0,t_1)$ the evolution of the system can be quite arbitrary and will be described
by a unitary evolution operator $U=U(t_1,t_0)$.\\

Next we will show that the suitably defined ``physical conditional probability" $p(j|i)$ is of modified doubly stochastic
type and hence the J-equation (\ref{P2b}) also holds in cases of physical relevance. Recall that the state of the system immediately after the
first measurement at time $t=t_0$ with outcome $i\in{\mathcal I}$ is assumed to be of the form (\ref{AS2a})
and that the time evolution between $t=t_0$ and $t=t_1$ is given by the
unitary evolution operator $U$. Hence according to the rules of quantum theory
\begin{equation}\label{G21}
  p(j|i)= \mbox{Tr}\left(
 \utilde{Q}_j\,U\,\frac{\utilde{P}_i}{d(i)}\,U^\ast  \right),\quad \mbox{for all } i\in{\mathcal I},\,j\in{\mathcal J}
  \;.
\end{equation}
For the sake of consistency we will show that $p(j|i)$ is a stochastic matrix. This follows by
\begin{equation}\label{G22}
  \sum_{j\in{\mathcal J}}p(j|i)
  \stackrel{(\ref{G21})}{=}
  \mbox{Tr}\left(\left( \sum_{j\in{\mathcal J}}\utilde{Q}_j \right)U\,\frac{\utilde{P}_i}{d(i)}\,U^\ast\right)
  \stackrel{(\ref{G8})}{=}
  \mbox{Tr}\left(U\,\frac{\utilde{P}_i}{d(i)}\,U^\ast\right)=\mbox{Tr}\left(\frac{\utilde{P}_i}{d(i)}\right)
   \stackrel{(\ref{G2})}{=}1
   \;,
\end{equation}
for all $i\in{\mathcal I}$. Moreover,
\begin{lemma}\label{LE2}
 The physical conditional probability (\ref{G21}) is of modified doubly stochastic type.
\end{lemma}

\noindent{\bf Proof}: According to the definition of ``modified doubly stochastic type" we have to confirm that (\ref{SM28}) holds:
\begin{equation}\label{G23}
  \sum_{i\in{\mathcal I}}p(j|i)\,d(i)
  \stackrel{(\ref{G21})}{=}
  \mbox{Tr}\left(\utilde{Q}_j\,U\,\left( \sum_{i\in{\mathcal I}}\utilde{P}_i\right)\,U^\ast\right)
  \stackrel{(\ref{G3})}{=}
 \mbox{Tr}\left(\utilde{Q}_j\,U\,U^\ast\right)= \mbox{Tr}\left(\utilde{Q}_j\right)
   \stackrel{(\ref{G7})}{=}D(j)
   \;,
\end{equation}
for all $j\in{\mathcal J}$.  \hfill$\Box$\\
Recall that certain ``hypothetical probabilities" $q(j)$ occur in Proposition \ref{Prop3} of section \ref{sec:SMM}.
For the quantum case we will always assume that these probabilities are of the following form:
\begin{equation}\label{G230}
q(j)=\mbox{Tr }\left( {\mathcal G}\left(\utilde{F}_1,\ldots,\utilde{F}_L\right)\,\utilde{Q}_j  \right)
    =D(j)\, {\mathcal G}\left(F_j^{\boldsymbol{\lambda}} \right)
,\mbox{ for all } j\in{\mathcal J}
\;,
\end{equation}
where the function $ {\mathcal G}$ is chosen to be the same as in (\ref{G5c}). We understand the ``standard case S" as including
the condition (\ref{G230}).

Lemma \ref{LE2} and Proposition \ref{Prop3} immediately entail the following theorem, referred to as claiming the general Jarzynski equation,
which will be only formulated for the standard case S:
\begin{theorem}\label{T1}
Let $\utilde{E}_1,\ldots,\utilde{E}_L$ be a family of mutually commuting self-adjoint operators with the spectral decomposition (\ref{G1})
satisfying (\ref{G2}), likewise $\utilde{F}_1,\ldots,\utilde{F}_L$ a second family satisfying (\ref{G6}) and (\ref{G7}). Further, let
$\rho={\mathcal G}\left( \utilde{E}_1,\ldots,\utilde{E}_L \right)$ be a density operator such that
\begin{equation}\label{G23a}
  p(i)= \mbox{Tr} \left( \rho\,\utilde{P}_i\right)>0
\end{equation}
holds for all $i\in{\mathcal I}$. Further, let $U$ be some unitary time evolution operator. Then the following holds
\begin{equation}\label{G24}
    \left\langle \frac{{\mathcal G}\left(F_j^{\boldsymbol{\lambda}} \right)}{{\mathcal G}\left(E_i^{\boldsymbol{\lambda}} \right)}\right\rangle =1
    \;,
  \end{equation}
  where the expectation value has been calculated by means of the physical probability function
  $p(i,j)=p(j|i)\,p(i)=\mbox{Tr}\left( \utilde{Q}_j\,U\,\frac{ \utilde{P}_i}{d(i)}\,U^\ast  \right)\,p(i)$.
\end{theorem}

We note in passing that the physical probabilities $p(i,j)$ can be written as
\begin{equation}\label{G25}
 p(i,j)=\mbox{Tr}\left( \rho\,F(i,j)\right)
 \;,
\end{equation}
where the positive operators
\begin{equation}\label{G26}
 F(i,j)\equiv \utilde{P}_i\,U^\ast\,\utilde{Q}_j\,U\,\utilde{P}_i\ge 0
\end{equation}
satisfy
\begin{equation}\label{G27}
  \sum_{(i,j)\in{\sf E}}F(i,j)=\sum_{i\in{\mathcal I}}\utilde{P}_i\,U^\ast\,\left( \sum_{j\in{\mathcal J}}\utilde{Q}_j\right)\,U\,\utilde{P}_i
  \stackrel{(\ref{G8})}{=}\sum_{i\in{\mathcal I}}\utilde{P}_i\stackrel{(\ref{G3})}{=}{\mathbbm 1}
\end{equation}
and hence constitute a ``positive operator valued measure" (POVM) $F:{\sf E}\rightarrow {\mathcal L}({\mathcal H})$.
Here ${\mathcal L}({\mathcal H})$ denotes the space of bounded, linear operators defined on ${\mathcal H}$.
(For a more general definition see \cite{BLPY16}; note that we use a simplified form of POVM adapted to countably infinite outcome spaces ${\sf E}$.)
Hence the various random variables defined on ${\sf E}$, including ``work", can be viewed as generalized observables in the sense of \cite{BLPY16},
albeit generally not ``sharp observables" (i.~e.~observables described by projection valued measures). This observation puts the
statement of \cite{TLH07}  ``work is not an observable" into perspective, see also \cite{RCP14}.

For the examples in the following subsections it suffices to identify the families $\utilde{E}_1,\ldots,\utilde{E}_L$
and $\utilde{F}_1,\ldots,\utilde{F}_L$ and the function ${\mathcal G}$. The conditions of Theorem \ref{T1} can be easily
verified by the reader; it remains to evaluate the general Jarzynski equation (\ref{G24}) for the various examples.
Moreover, since in all examples the convex exponential function is involved we may invoke Jensen's inequality and obtain
special relations that may be viewed as manifestations of the $2^{nd}$ law for non-equilibrium case S scenarios, but have to be
distinguished from the case R statement of non-decreasing modified Shannon entropy in subsection \ref{sec:SG}.

%%%%%%%%%%%%%%%%%%%%%%%%%%%%%%%%%%%%%%%%%%%%%%%%%%%%%%%%%%%%%%%%%%%%%%%%%%%%%%%%%%%%%%%%%%%%%%%%%%%%%%%%%%%%%%%%%%%%%%%%%%%%%%%
\subsection{Systems in local canonical equilibrium}\label{sec:L}
%%%%%%%%%%%%%%%%%%%%%%%%%%%%%%%%%%%%%%%%%%%%%%%%%%%%%%%%%%%%%%%%%%%%%%%%%%%%%%%%%%%%%%%%%%%%%%%%%%%%%%%%%%%%%%%%%%%%%%%%%%%%%%%

We assume that the quantum system consists of $N$ subsystems and consequently ${\mathcal H}=\bigotimes_{\mu=1}^N {\mathcal H}_\mu$.
For each subsystem we assume a, possibly time-dependent, Hamiltonian $H_\mu(t)$, where the lift to the total Hilbert space by means
of suitable tensor products with identity operators will be tacitly understood.
Its spectral composition will be written as
\begin{equation}\label{L0}
  H_\mu(t)=\sum_{i_\mu\in{\mathcal I}_\mu}E_{i_\mu}^{(\mu)}(t)\,P_{i_\mu}(t)
  \;.
\end{equation}
Then we set $L=N$, ${\mathcal I}={\mathcal J}= {\mathcal I}_1\times\ldots\times {\mathcal I}_N$ and
\begin{equation}\label{L1}
 \utilde{E}_\mu =H_\mu(t_0),\quad \mbox{and}\quad  \utilde{F}_\mu =H_\mu(t_1),\quad \mbox{for } \mu=1,\ldots N
 \;.
\end{equation}
These Hamiltonians are not necessarily connected with the unitary time evolution operator $U=U(t_1,t_0)$.
Further, we choose
\begin{equation}\label{L2}
  \rho={\mathcal G}\left(H_1(t_0),\ldots,H_N(t_0)\right)=\prod_{\mu=1}^{N}
  \left(\mbox{Tr}\,\exp\left(-\beta_\mu\,H_\mu(t_0) \right) \right)^{-1}\,\exp\left(-\beta_\mu\,H_\mu(t_0) \right)
  \;,
\end{equation}
with the usual interpretation of the parameters $\beta_\mu>0$ as the inverse temperatures of the subsystems.
The generalized Jarzynski equation (\ref{G24}) then assumes the form
\begin{equation}\label{L3}
 \left\langle \exp\left(-\sum_{\mu=1}^{N}\beta_\mu\,\left(w_\mu-\Delta F_\mu\right) \right)\right\rangle =1
 \;,
\end{equation}
where
\begin{eqnarray}
\label{L4a}
  w_\mu(i,j) &\equiv& E_{j_\mu}^{(\mu)}(t_1)-E_{i_\mu}^{(\mu)}(t_0), \\
  \label{L4b}
  Z_\mu(t) &\equiv& \mbox{Tr}\left(e^{-\beta_\mu\,H_\mu(t)} \right) \equiv e^{-\beta_\mu\,F_\mu(t)},\\
  \label{L4c}
   \Delta F_\mu &\equiv&  F_\mu(t_1)-F_\mu(t_0),\quad \mbox{for all } \mu=1,\ldots,N
   \;.
\end{eqnarray}
Since $\exp$ is convex Jensen's inequality yields $e^{\langle x\rangle}\le \left\langle e^x\right\rangle$, and hence (\ref{L3}) implies
\begin{equation}\label{L5}
  \exp\left\langle -\sum_{\mu=1}^{N}\beta_\mu\,\left(w_\mu-\Delta F_\mu\right)\right\rangle\le 1
  \;,
\end{equation}
or, equivalently,
\begin{equation}\label{L6}
 \sum_{\mu=1}^{N}\beta_\mu\,\left(\left\langle w_\mu\right\rangle-\Delta F_\mu\right)\ge 0
  \;.
\end{equation}
Note that the l.~h.~s.~of (\ref{L6}) has the form of a sum of entropy changes in the quasi-static limit and hence can be
viewed as a manifestation of the $2^{nd}$ law for the present non-equilibrium scenario.
For similar results see \cite{T00} and \cite{CPF17}.

%%%%%%%%%%%%%%%%%%%%%%%%%%%%%%%%%%%%%%%%%%%%%%%%%%%%%%%%%%%%%%%%%%%%%%%%%%%%%%%%%%%%%%%%%%%%%%%%%%%%%%%%%%%%%%%%%%%%%%%%%%%%%%%
\subsection{Systems in micro-canonical equilibrium}\label{sec:M}
%%%%%%%%%%%%%%%%%%%%%%%%%%%%%%%%%%%%%%%%%%%%%%%%%%%%%%%%%%%%%%%%%%%%%%%%%%%%%%%%%%%%%%%%%%%%%%%%%%%%%%%%%%%%%%%%%%%%%%%%%%%%%%%

We choose $L=1$ and a one-parameter family of Hamiltonians $H(t)$ with spectral decomposition
\begin{equation}\label{M1}
  H(t)=\sum_{i\in{\mathcal I}}E_i(t)\,P_i(t)
  \;,
\end{equation}
and a representation of the micro-canonical ensemble in the form
\begin{equation}\label{M2}
 \rho={\mathcal G}(H(t_0))\equiv \frac{1}{W(t_0)}\exp\left(-\left(\frac{E-H(t_0)}{w}\right)^2 \right)
 \;,
\end{equation}
where
\begin{equation}\label{M3}
 W(t)\equiv \mbox{Tr }\exp\left[-\left(\frac{E-H(t)}{w}\right)^2 \right]\equiv e^{-f(t)}
 \;,
\end{equation}
and $E,w>0$ are parameters.
The generalized Jarzynski equation (\ref{G24}) then assumes the form
\begin{equation}\label{M4}
 \left\langle \exp\left[-\left(\frac{E-E_j(t_1)}{w} \right)^2+\Delta f+\left(\frac{E-E_i(t_0)}{w} \right)^2 \right]\right\rangle =1
 \;,
\end{equation}
where $\Delta f\equiv f(t_1)-f(t_0)$. Application of Jensen's inequality analogous to that in Section \ref{sec:L} yields
\begin{equation}\label{M4}
 \left\langle \left(\frac{E-E_j(t_1)}{w} \right)^2-\Delta f-\left(\frac{E-E_i(t_0)}{w} \right)^2 \right\rangle\ge 0
 \;.
\end{equation}

The generalization to systems in local micro-canonical equilibrium analogous to the case treated in Section \ref{sec:L}
is straightforward and need not be given here in detail.

%%%%%%%%%%%%%%%%%%%%%%%%%%%%%%%%%%%%%%%%%%%%%%%%%%%%%%%%%%%%%%%%%%%%%%%%%%%%%%%%%%%%%%%%%%%%%%%%%%%%%%%%%%%%%%%%%%%%%%%%%%%%%%%
\subsection{Systems in grand canonical equilibrium}\label{sec:R}
%%%%%%%%%%%%%%%%%%%%%%%%%%%%%%%%%%%%%%%%%%%%%%%%%%%%%%%%%%%%%%%%%%%%%%%%%%%%%%%%%%%%%%%%%%%%%%%%%%%%%%%%%%%%%%%%%%%%%%%%%%%%%%%

The Hilbert space of the system is chosen as the bosonic or fermionic Fock space over the one-particle Hilbert space ${\mathcal H}$:
\begin{equation}\label{R1}
 {\mathcal F}_\pm(\mathcal H)=\bigoplus_{n=0}^\infty {\mathcal S}_\pm {\mathcal H}^{\otimes n}
 \;,
\end{equation}
where ${\mathcal H}^{\otimes n}$ denotes the $n$-fold tensor product and ${\mathcal S}_\pm$ the projector onto
the totally symmetric (+) part or the totally anti-symmetric (-) part of of ${\mathcal H}^{\otimes n}$.
We choose $L=2$ and $\utilde{E}_1=H(t_0)$, where $H(t)$ is the canonical lift of a time-dependent one-particle Hamiltonian $H_1(t)$
to  ${\mathcal F}_\pm(\mathcal H)$. Further, we choose $\utilde{E}_2=\utilde{N}$, the particle number operator in  ${\mathcal F}_\pm(\mathcal H)$.
By definition, $\utilde{E}_1$ and $\utilde{E}_2$ commute. Let the respective spectral decompositions with a common system of eigenprojections
be written as
\begin{equation}\label{R2}
 H(t)=\sum_{i\in{\mathcal I}}E_i(t)\,P_i(t)
 \;,
\end{equation}
and
\begin{equation}\label{R3}
 \utilde{N}=\sum_{i\in{\mathcal I}}N_i\,P_i(t)
 \;.
 \end{equation}
Moreover, we set
\begin{equation}\label{R4}
  \rho={\mathcal G}\left(H(t_0),\utilde{N}\right) \equiv
  \exp\left[ \beta\left(\Omega(t_0)+\mu\,\utilde{N}-H(t_0)\right)\right]
  \;,
\end{equation}
where
\begin{equation}\label{R5}
 \exp\left(-\beta\, \Omega(t)\right)\equiv \mbox{Tr}\exp\left[\beta\left(\mu\,\utilde{N}-H(t)\right)\right]
 \;,
\end{equation}
and $\beta,\mu,\Omega$ have the usual physical interpretation as inverse temperature, chemical potential and grand potential, respectively.

The generalized Jarzynski equation (\ref{G24}) then assumes the form
\begin{equation}\label{R6}
 \left\langle \exp\left[-\beta \left(\left(E_j(t_1)-E_i(t_0) \right)-\mu\left(N_j-N_i \right)-\Delta \Omega \right)\right]\right\rangle =1
 \;,
\end{equation}
where $\Delta \Omega \equiv \Omega(t_1)-\Omega(t_0)$. Application of Jensen's inequality analogous to that in Section \ref{sec:L} yields
\begin{equation}\label{R7}
 \beta \left\langle\left(E_j(t_1)-E_i(t_0) \right\rangle-\mu\left\langle N_j-N_i \right\rangle-\Delta \Omega \right)\ge 0
 \;.
\end{equation}

The generalization to systems in local grand canonical equilibrium analogous to the case treated in Section \ref{sec:L}
is straightforward and need not be given here in detail. For similar results see also \cite{SS07} - \cite{YKT12}.

%%%%%%%%%%%%%%%%%%%%%%%%%%%%%%%%%%%%%%%%%%%%%%%%%%%%%%%%%%%%%%%%%%%%%%%%%%%%%%%%%%%%%%%%%%%%%%%%%%%%%%%%%%%%%%%%%%%%%%%%%%%%%%%
\subsection{Application to periodic thermodynamics}\label{sec:PT}
%%%%%%%%%%%%%%%%%%%%%%%%%%%%%%%%%%%%%%%%%%%%%%%%%%%%%%%%%%%%%%%%%%%%%%%%%%%%%%%%%%%%%%%%%%%%%%%%%%%%%%%%%%%%%%%%%%%%%%%%%%%%%%%

Analogously to Section \ref{sec:L} we consider two systems (i.~e.~$N=2$) and assume that the first system
is periodically driven with a Hamiltonian $K_1(t)$ satisfying $K_1(t+T)=K_1(t)$.
We have chosen the letter ``K" since we will have to distinguish between the Hamiltonian and the (quasi) energy operator $H_1(t)$
and want to conform, as far as possible, with the notation introduced in the preceding sections.
According to
Floquet theory the general solution of the corresponding Schr\"odinger equation will be of the form
\begin{equation}\label{PT1}
 \psi(t)=\sum_{i\in{\mathcal I}_1}a_i\,u_i(t)\,e^{-{\sf i}\,\epsilon_i\,t}
\;,
\end{equation}
with time-independent coefficients $a_i$. Here the $\epsilon_i$ denote the quasi-energies, unique up to integer multiples of
$\omega \equiv \frac{2\pi}{T}$, and the $u_i(t)$ are $T$-periodic functions of $t$. We assume a pure point spectrum of the quasi-energies
and accordingly ${\mathcal I}_1$ will be a countably infinite or possibly finite index set.

Upon choosing a selection of quasi-energies from their equivalence classes
we may define a quasi-energy operator
\begin{equation}\label{PT2}
  H_1(t)= \sum_{i\in{\mathcal I}_1}\epsilon_i\,\utilde{P}^{(1)}_i(t)\equiv  \sum_{i\in{\mathcal I}_1}\epsilon_i\,\left| u_i(t)\rangle\langle u_i(t)\right|
  \;.
\end{equation}
Hence $\mbox{Tr }\utilde{P}^{(1)}_i(t)=1$ for all $t$
and
\begin{equation}\label{PT2a}
  \sum_{i\in{\mathcal I}_1}\utilde{P}^{(1)}_i(t)={\mathbbm 1}
  \;.
\end{equation}

The first system is coupled to a heat bath with Hamiltonian
\begin{equation}\label{PT3}
  H_2=\sum_{n\in{\mathcal N}}E_n\,\utilde{P}^{(2)}_n
  \;,
\end{equation}
where the $\utilde{P}^{(2)}_n$ are assumed to be finite-dimensional projectors with dimension (degeneracy) $d(n)\mbox{Tr }\utilde{P}^{(2)}_n$.
Without loss of generality we also assume a pure point spectrum of the heat bath corresponding to a countably infinite index set ${\mathcal N}$.
The outcome sets introduced in Section \ref{sec:SM} can be chosen as ${\mathcal I}={\mathcal J}={\mathcal I}_1\times {\mathcal N}$.
Note the completeness relation
\begin{equation}\label{PT4}
 \sum_{n\in{\mathcal N}}\utilde{P}^{(2)}_n={\mathbbm 1}
  \;.
\end{equation}

The total Hamilton operator of the system plus bath will be written as
\begin{equation}\label{PT5}
K(t)=K_1(t)\otimes {\mathbbm 1}_2+{\mathbbm 1}_1\otimes H_2 +H_{12}
 \;,
\end{equation}
with some self-adjoint operator $H_{12}$ defined on the total Hilbert space ${\mathcal H}={\mathcal H}_1\otimes{\mathcal H}_2$
describing the system-bath interaction. It is assumed to be valid for $t<t_0$. Strictly speaking, the form of $K(t)$ is
irrelevant for the Jarzynski equation to be formulated below. Its only purpose is to motivate the following
assumptions about the state of the total system at the time $t=t_0$.

We assume that for times $t<t_0$ the heat bath will be in a thermal equilibrium state
\begin{equation}\label{PT6}
 \rho_2=\frac{1}{Z_2}\,e^{-\beta H_2}
 \;,
 \end{equation}
 where, as usual, $\beta$ is the inverse temperature and the heat bath partition function is
 \begin{equation}\label{PT7}
  Z_2=\mbox{Tr}\left(e^{-\beta H_2}\right)= \sum_{n\in{\mathcal N}}e^{-\beta E_n}\,d(n)
  \;.
 \end{equation}

The crucial assumption of this subsection will be that also the system assumes, for times $t<t_0$,
a quasi-stationary distribution $\rho_1(t)$ of Floquet states that will be of Boltzmann type
with an inverse quasi-temperature $\vartheta$, namely
\begin{equation}\label{PT8}
 \rho_1(t)=\frac{1}{Z_1}\,e^{-\vartheta H_1(t)}
 \;,
\end{equation}
and the corresponding time-independent quasi-partition function reads
\begin{equation}\label{PT9}
 Z_1=\mbox{Tr}\left( e^{-\vartheta H_1(t)}\right)=\sum_{i\in{\mathcal I}_1}e^{-\vartheta\,\epsilon_i}
 \;.
\end{equation}
W.~r.~t.~the conditions of Theorem \ref{T1} we thus may write the initial state as
\begin{equation}\label{PT10}
\rho=\rho_1(t_0)\otimes \rho_2= {\mathcal G}\left( H_1(t_0),H_2\right)=
\left[\mbox{Tr}\left( e^{-\vartheta\,H_1(t_0)}\right) \mbox{Tr}\left( e^{-\beta\,H_2}\right)\right]^{-1}
\exp\left(-\vartheta\,H_1(t_0)-\beta\,H_2 \right)
\;.
\end{equation}

Whereas the general existence of a quasi-stationary distribution has been made plausible in the literature \cite{BHP00},
the more restrictive assumption of a quasi Boltzmann distribution has only be demonstrated for four kinds of systems:\\
\begin{itemize}
  \item For the particular case of a linearly forced
harmonic oscillator the authors of \cite{BHP00} have shown that the
Floquet-state distribution remains a Boltzmann distribution
with the temperature of the heat bath, i.~e.~$\vartheta=\beta$, see also \cite{LH14}.
\item Similarly, the parametrically driven harmonic oscillator assumes a quasi-stationary state with a quasi-temperature
that is, however, generally different from the bath temperature, see \cite{DFH19}.
  \item A spin $s$  exposed
to both a static magnetic field and an oscillating,
circularly polarized magnetic field applied perpendicular
to the static one, as in the classic Rabi set-up,
and coupled to a thermal bath of harmonic oscillators is shown to approach a quasi Boltzmann distribution in \cite{SSH19},
  \item and finally every quasi-stationary distribution of Floquet states of a two-level system, see \cite{LH14}, can be trivially viewed
  as a quasi Boltzmann distribution.
\end{itemize}

As in the section \ref{sec:Q} we will assume that at the times $t=t_0$ and $t=t_1$ there will be performed measurements
of the observables corresponding to the commuting (quasi) energy operators $H_1(t)$ and $H_2$. The interaction
between the system and the heat bath in the time interval $(t_0,t_1)$ can be quite arbitrary and will be described by a Hamiltonian $\tilde{H}(t)$.
It follows that all mathematical assumptions necessary to prove the general Jarzynski equation (\ref{G24}) are satisfied.
But note the following difference: Typically the general Jarzynski equation holds
in a situation of local thermal equilibrium at the initial time $t=t_0$. In this section we will rather apply Theorem \ref{T1} to a situation
of a quasi-stationary distribution of Floquet states of a periodically driven system in contact with a heat bath.
This situation may be far from local thermal equilibrium.

Analogously to Section \ref{sec:L} we will set $\beta_1=\vartheta$, the inverse quasi-temperature, whereas
$\beta_2=\beta$ is the ordinary temperature of the heat bath. Consequently, we will rewrite $w_1$ as the ``change of quasi-energy  $e$"
and $w_2$ as the heat $q$ absorbed by the heat bath.  Strictly speaking, this would exclude a time-dependent Hamiltonian for the
heat bath since otherwise $w_2$ could also be composed of both, heat and work. Nevertheless, we will stick to this more intuitive notation.

A noted above, both partition functions $Z_1$ and $Z_2$ are time-independent. Hence  (\ref{G24}) simplifies to
\begin{equation}\label{PT11}
  \left\langle \exp\left( -\vartheta\,e-\beta\,q\right)\right\rangle =1
  \;.
\end{equation}
The inequality derived by means of Jensen's inequality analogous to (\ref{L6}) hence will read
\begin{equation}\label{PT12}
  \vartheta\,\left\langle e\right\rangle+\beta\,\left\langle q\right\rangle\ge 0
  \;,
\end{equation}
and can again be viewed as a manifestation of the $2^{nd}$ law for periodic thermodynamics.

%%%%%%%%%%%%%%%%%%%%%%%%%%%%%%%%%%%%%%%%%%%%%%%%%%%%%%%%%%%%%%%%%%%%%%%%%%%%%%%%%%%%%%%%%%%%%%%%%%%%%%%%%%%%%%%%%%%%%%%%%%%%%%%
\section{Further applications to quantum theory}\label{sec:S}
%%%%%%%%%%%%%%%%%%%%%%%%%%%%%%%%%%%%%%%%%%%%%%%%%%%%%%%%%%%%%%%%%%%%%%%%%%%%%%%%%%%%%%%%%%%%%%%%%%%%%%%%%%%%%%%%%%%%%%%%%%%%%%%
\subsection{A Second Law-like statement for the non-standard case}\label{sec:SG}
%%%%%%%%%%%%%%%%%%%%%%%%%%%%%%%%%%%%%%%%%%%%%%%%%%%%%%%%%%%%%%%%%%%%%%%%%%%%%%%%%%%%%%%%%%%%%%%%%%%%%%%%%%%%%%%%%%%%%%%%%%%%%%%

In the preceding sections we have formulated a number of $2^{nd}$ law-like statements, namely (\ref{L6}), (\ref{M4}), (\ref{R7}), and (\ref{PT12}),
that follow from the respective Jarzynski equations in the standard case S. However, these statements are not special cases of the
``Pauli-type" inequalities (\ref{SM18}) and (\ref{SM30}) since these are based on the assumption $q(j)=\hat{p}(j)$ for all $j\in{\mathcal J}$
(case R)
and hence do not belong to the standard case that is characterized by (\ref{G230}). In view of the fundamental significance of the Second Law
it will be in order to add a few remarks on the realization of (\ref{SM18}) and (\ref{SM30}) in quantum mechanics.

First, we will reformulate (\ref{SM30}) in the context of quantum theory:
\begin{theorem}\label{T2}
We assume the notations and general conditions of Section \ref{sec:Q}, in particular the Assumptions $1$ and $2$.
It follows that the $5$-tuple
 $({\mathcal I}, {\mathcal J}, p, d, D)$
 will be a modified statistical model of sequential measurements ($mST^2$) where the physical probability function
 $p:{\mathcal I}\times{\mathcal J}\rightarrow{\mathbbm R}$ is given by
 \begin{equation}\label{S1}
   p(i,j)=\mbox{Tr}\left( \utilde{Q}_j\,U\, \utilde{P}_i\,\rho\, \utilde{P}_i\,U^\ast\right)=p(j|i)\,p(i)
   \;,
 \end{equation}
 and the second marginal probabilities are defined by
 \begin{equation}\label{S2}
   \hat{p}_j=\sum_{i}p(i,j)
   \;.
 \end{equation}
 Then the following holds:
 \begin{equation}\label{S3}
S'(\hat{p})=-\sum_j  \hat{p}_j\,\log \frac{\hat{p}_j}{D(j)}\ge S'(p)= -\sum_i p(i)\,\log\frac{ p(i) }{d(i)}
  \;.
 \end{equation}
\end{theorem}

This statement is certainly not new but has a couple of forerunners albeit formulated in different frameworks
\cite{GS14}, \cite{P70}, and \cite{P28}.
We note that the modified Shannon entropy $ S'(p)$ can be identified with the von Neumann entropy $-\mbox{Tr}\left(\rho_1\,\log\, \rho_1\right)$
of the mixed state $\rho_1$ after the first measurement according to (\ref{G5b}). Indeed,
\begin{equation}\label{S4}
  \rho_1\stackrel{(\ref{G5b})}{=}\sum_i p(i)\,\rho_i\stackrel{(\ref{AS2a})}{=}\sum_i \frac{p(i)}{d(i)}\,P_i
\end{equation}
implies
\begin{equation}\label{S5}
 - \mbox{Tr}\left(\rho_1\,\log\, \rho_1\right)= -\mbox{Tr}\left(\sum_i \frac{p(i)}{d(i)}\,\log\frac{p(i)}{d(i)}\, \,P_i\right)
  =-\sum_i p(i) \log\frac{p(i)}{d(i)}=S'(p)
  \;.
\end{equation}
For the modified Shannon entropy $S'(\hat{p})$ this identification is not possible in general. Even if we additionally assume that the second
measurement will be of L\"uders type it is not clear whether an assumption analogous to (\ref{AS2a}) would hold. Below we will
consider a simplified scenario where this identification is nevertheless possible.

\begin{figure}[t]
\centering
\includegraphics[width=0.7\linewidth]{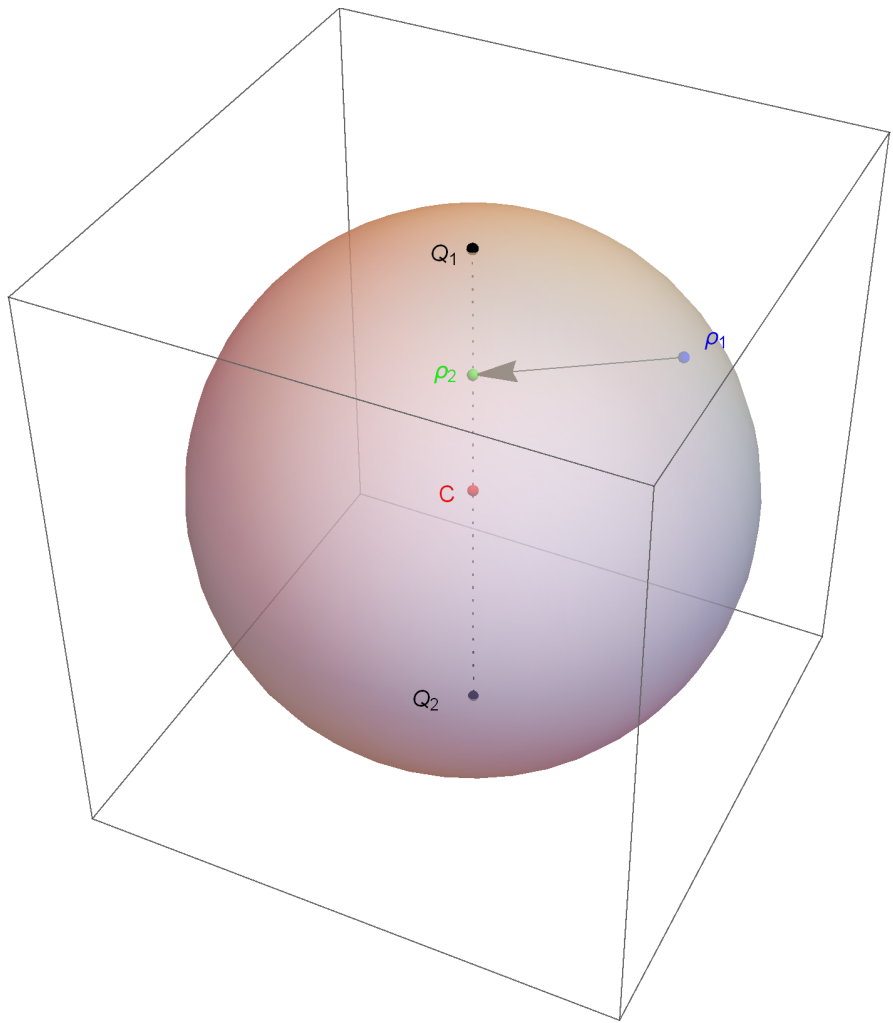}
\caption{Geometric interpretation of the L\"uders operation $\rho_1\mapsto \rho_2$ as a projection
of points inside the Bloch sphere onto the line joining two orthogonal projections $Q_1$ and $Q_2$. }
\label{FIGBLOCH}
\end{figure}

Next we note that the $2^{nd}$ law-like statement (\ref{S3}) holds for closed systems irrespective of their size and is in this respect
more general than the usual formulations of the Second Law for large systems including small systems coupled to a heat bath.
Moreover, (\ref{S3}) is not restricted to sequential energy measurements and, e.~g., would also hold for (discretized) position
measurements thereby describing the spreading of wave-packets, see the example of the following subsection \ref{sec:SA}.
In this context it might be instructive to discuss the well-known
Umkehreinwand (reversibility paradox) of Loschmidt. There exist solutions $\psi(t)$ of, say, the $1$-particle Schr\"odinger equations that are time-reflections of
spreading wave-packets and hence concentrate on smaller and smaller regions. These solutions do not lead to a violation of (\ref{S3})
since after the first measurement this special solution $\psi(0)$  is transformed into a mixed state $\rho_1$ that again will spread with increasing time.
The delicate phase relations of $\psi(0)$ needed for the inverse spreading are destroyed by the first measurement.

Similarly, the related Wiederkehreinwand (recurrence paradox) of Poincar\'{e} and Zermelo
that would be particular serious for small systems with short recurrence times can be rebutted.
It may happen that the modified Shannon entropy $S'(\hat{p})$ will be a periodic function $s(t)$ of the time difference $t\equiv t_1-t_0$ between the two measurements
but this does not injure the validity of (\ref{S3}). The reason is simply that the latter inequality reads $s(t)\ge S'(p)$ and {\it not}
$s(t_a)\ge s(t_b)$ for all $t_a>t_b$. Physically speaking, the Wiederkehreinwand does not apply since in quantum mechanics the
entropy difference is not a definite quantity defined for all times $t$ but rather should be construed as the mean value of entropy differences over many measurements of a pair of observables performed at a fixed time difference $t$.

If (\ref{S3}) is a fundamental inequality that is valid for a large class of sequential measurements it should have a simple interpretation
that we are now going to explain. To this end we generalize our considerations to a finite number of $L$  sequential L\"uders measurements but restricted to
the case of a finite $n$-dimensional Hilbert space ${\mathcal H}$ and non-degenerate projections $P_i$, $Q_j$. This corresponds to the subsection
\ref{sec:SMS} dealing with the ``simple case". In particular the Assumption $2$, see (\ref{AS2a}), will be satisfied for the corresponding state
before each measurement. Consider first the simplest case of an $n=2$-dimensional Hilbert space where all mixed states correspond to the points of a unit ball
with center $C\cong \frac{1}{2}{\mathbbm 1}$. The boundary of the unit ball is usually denoted as the ``Bloch sphere".
Two orthogonal projections $P_1$ and $P_2$ are represented by antipodal pairs of points of the Bloch sphere and the first L\"uders operation
$\rho\mapsto \rho_1$ is just the projection onto the line joining $P_1$ and $P_2$. Upon this projection the distance of the state to the center $C$
decreases (or remains constant). This distance can be expressed in terms of the scalar product
$(A,B)\mapsto \mbox{Tr} A^\ast\,B$ for $A,B\in{\mathcal L}({\mathcal H})$.
The unitary time evolution between the first and the second measurement corresponds to a rotation of the Bloch sphere and can be discarded
as far as only geometric relations are considered. Then the second measurement with orthogonal projections $Q_1$ and $Q_2$ again yields
a projection that maps $\rho_1$ onto, say, $\rho_2$ and further decreases the distance to the center $C$. In this way the $L$ sequential L\"uders
measurements yield a sequence $\rho,\,\rho_1,\,\rho_2,\ldots,\rho_L$ of mixed states represented by points inside the Bloch sphere with non-increasing
distance to the center $C$, see Figure \ref{FIGBLOCH}.

Before we connect this geometric picture to the $2^{nd}$ law-like statement (\ref{S3}) we will sketch the generalization to finite $n>2$
although the corresponding geometry cannot be visualized in a likewise simple manner. The unit ball in the case of $n=2$ has to be replaced by the
convex set $K$ of mixed states such that the pure states are the extremal points of $K$. The center $C$ now corresponds to the mixed state
$C\cong \frac{1}{n}{\mathbbm 1}$.  A family of $n$ mutually orthogonal $1$-dimensional projections $P_i$ spans an $n$-simplex $\Sigma$ of extremal points
of $K$ consisting of all mixed states that commute with all $P_i,\,i=1,\ldots,n$. The center $C$ is always contained in the simplex $\Sigma$.
The L\"uders operation $\rho\mapsto \rho_1$ is the projection onto the affine subspace spanned by the $P_i,\,i=1,\ldots,n$.
Again, by subsequent projections we obtain a sequence $\rho,\,\rho_1,\,\rho_2,\ldots,\rho_L$
of mixed states such that the distance to the center $C$ is non-increasing.

The connection to the  $2^{nd}$ law-like statement (\ref{S3}) will be only discussed for the special case of $n=2$. In the common
eigenbasis of $\utilde{Q}_1$, $\utilde{Q}_2$ the L\"uders operation $\rho_1\mapsto \rho_2$  assumes the form
\begin{equation}\label{S6}
  \left(
  \begin{array}{cc}
    p&\lambda\,e^{{\sf i}\alpha}\\
  \lambda\,e^{-{\sf i}\alpha} & 1-p
  \end{array}
  \right)
  \mapsto\left(
  \begin{array}{cc}
    p&0\\
   0 & 1-p
  \end{array}
  \right)
  \;,
\end{equation}
for some $\lambda\in[0,\sqrt{p(1-p)}]$ depending on $\rho_1$. We may regard $\lambda$ as a parameter and denote the matrix at the l.~h.~s.~of (\ref{S6})
by $\rho(\lambda)$. This matrix function parametrizes the projection onto the line joining $\utilde{Q}_1$ and $\utilde{Q}_2$ by a smooth curve (line) inside the Bloch sphere.
$\rho(\lambda)$ has the eigenvalues
\begin{equation}\label{S7}
 p_{1,2}(\lambda)=\frac{1}{2}\pm \sqrt{\lambda^2+\frac{1}{4}-p(1-p)}
 \;.
\end{equation}
The squared distance to the center $C\cong \frac{1}{2}\mathbbm{1}$ can be calculated in the eigenbasis of $\rho(\lambda)$
with the result
\begin{equation}\label{S8}
\left\|\rho(\lambda)- \frac{1}{2}\mathbbm{1}\right\|^2=\left(p_1-\frac{1}{2} \right)^2+\left(p_2-\frac{1}{2} \right)^2=2 \lambda ^2-2 (1-p) p+\frac{1}{2}
 \;.
\end{equation}
It is obviously a monotonically increasing function of $\lambda$. The von Neumann entropy $S(\rho(\lambda))$ is given by
\begin{equation}\label{S9}
  S(\rho(\lambda))=-p_1(\lambda)\log p_1(\lambda)-p_2(\lambda)\log p_2(\lambda)
  \;.
\end{equation}
Inserting (\ref{S7}) and after some manipulations we obtain
\begin{equation}\label{S10}
  \frac{\partial S(\rho(\lambda))}{\partial\lambda}=-\frac{4 \lambda  \tanh ^{-1}\left(\sqrt{4 \lambda ^2+(1-2 p)^2}\right)}{\sqrt{4
   \lambda ^2+(1-2 p)^2}}
   \;.
\end{equation}
This derivative is negative for $0<\lambda\le \sqrt{p(1-p)}$ and vanishes only for $\lambda=0$. Hence $S(\rho(\lambda))$  is a monotonically
decreasing function of $\lambda$. Summarizing, if restricted to the curve $\rho(\lambda)$, the von Neumann entropy is a
monotonically decreasing function of the distance to the center $C$ and the monotonic decrease of the distance during the process
of sequential measurements results in the corresponding increase of the von Neumann entropy.

For the general case of $n=\dim {\mathcal H}>2$ the above result that the von Neumann entropy is a
monotonically decreasing function of the distance to the center $C$ along the curve $\rho(\lambda)$ still holds,
but we have not found a proof independent of Theorem \ref{T2}.

%%%%%%%%%%%%%%%%%%%%%%%%%%%%%%%%%%%%%%%%%%%%%%%%%%%%%%%%%%%%%%%%%%%%%%%%%%%%%%%%%%%%%%%%%%%%%%%%%%%%%%%%%%%%%%%%%%%%%%%%%%%%%%%
\subsection{An analytically solvable example}\label{sec:SA}
%%%%%%%%%%%%%%%%%%%%%%%%%%%%%%%%%%%%%%%%%%%%%%%%%%%%%%%%%%%%%%%%%%%%%%%%%%%%%%%%%%%%%%%%%%%%%%%%%%%%%%%%%%%%%%%%%%%%%%%%%%%%%%%

As a non-trivial example we consider a particle in one dimension and a free time evolution between the two measurements described by the
Schr\"odinger equation
\begin{equation}\label{SA1}
  {\sf i}\,\hbar \frac{\partial}{\partial t}\psi(x,t)=-\frac{\hbar^2}{2 m}\frac{\partial^2}{\partial x^2}\psi(x,t)
\end{equation}
with self-explaining notation. The two measurements are unsharp position-momentum measurements. More specifically,
the projections $\utilde{P}_i,\,i\in{\mathcal I}$ considered in Section \ref{sec:Q} are one-dimensional and of the form  $\utilde{P}_i=\left|i\rangle\langle i \right|$ where
\begin{equation}\label{SA2}
  |i\rangle=|\nu,n\rangle\equiv \chi_{[\nu\Delta,(\nu+1)\Delta]}\, \frac{1}{\sqrt{\Delta}}\,\exp\left(\frac{2\pi{\sf i}\,n\,x}{\Delta} \right)
  \;,\; \nu,n\in{\mathbbm Z}
  \;.
\end{equation}
Here $\chi_{[\nu\Delta,(\nu+1)\Delta]}$ denotes the characteristic function of the interval $[\nu\Delta,(\nu+1)\Delta]$.
Thus the first measurement is a discretized joint measurement of position $q_\nu$ and momentum $p_n=\frac{2\,\pi\,\hbar\,n}{\Delta}$.
Analogous definitions hold for the second measurement with projections $\utilde{Q}_j=|j\rangle\langle j|$ and $|j\rangle=|\mu,m\rangle$
such that ${\mathcal I}={\mathcal J}={\mathbbm Z}^2$.

We choose the physical units such that $\Delta=m=\hbar=1$ and an initial pure state given by the Gaussian
\begin{equation}\label{SA3}
 \psi(x)=\frac{1}{\pi^{1/4}\sigma}\exp\left(- \frac{x^2}{2\sigma^2}\right)
 \;.
\end{equation}
After the first (L\"uders) measurement the particle is in one of the pure states $|\nu,n\rangle$ with probability
\begin{equation}\label{SA3a}
 p(i)=p(\nu,n)=\left|\langle \nu,n|\psi\rangle \right|^2
 \;,
\end{equation}
where
\begin{equation}\label{SA4}
 \langle \nu,n|\psi\rangle =\int_{\nu}^{\nu+1}\psi(x) \, \exp\left(-2\pi{\sf i}\,n\,x\right)\,dx=
  \frac{\sqrt[4]{\pi } }{\sqrt{2}}\sqrt{\sigma } e^{-2 \pi ^2 n^2 \sigma ^2}
  \left(\text{erf}\left(\frac{\nu +1+2  \pi {\sf i}  n \sigma ^2}{\sqrt{2} \sigma}\right)-
  \text{erf}\left(\frac{\nu +2  \pi  {\sf i} n \sigma ^2}{\sqrt{2} \sigma   }\right)\right)
   \;.
\end{equation}
After the time $t=t_1-t_0$ the state $|\nu,n\rangle$  evolves into $|\nu,n,t\rangle$.
To calculate the latter we have integrated the free propagator over one unit interval and thereafter performed a Galilean boost
with the result
\begin{equation}\label{SA5}
  |\nu,n,t\rangle =\frac{{\sf i}}{2} e^{-2  \pi {\sf i} n (\pi  n t-x)} \left(
     \text{erfi}\left(\frac{1+{\sf i}} {2\sqrt{t}}
   (\nu +2 \pi  n t-x)\right)
   -
    \text{erfi}\left(\frac{1+{\sf i}} {2\sqrt{t}}
   (\nu+1 +2 \pi  n t-x)\right)\right)
   \;.
\end{equation}
Here $\text{erfi}(z)$ denotes the imaginary error function $\frac{\text{erf}({\sf i}z)}{{\sf i}}$. The second (L\"uders)
measurement yields the result $j=(\mu,m)$ with conditional probability
\begin{equation}\label{SA6}
  p(j|i)=p(\mu,m|\nu,n)=\left| \langle \mu,m|\nu,n,t\rangle\right|^2
  \;.
\end{equation}
If $m\neq n$ the corresponding amplitudes are obtained as
\begin{eqnarray}\nonumber
   \langle \mu,m|\nu,n,t\rangle&=&\frac{e^{-4 {\sf i} \pi^2 t \left(m^2-2 m n+2 n^2\right)}}{4 \pi  (m-n)}
   \left(
   e^{2 {\sf i} \pi^2 t (m-2 n)^2} (-2 R(m,\mu ,\nu )+R(m,\mu ,\nu +1)+R(m,\mu +1,\nu ))\right.\\
   \label{SA7}
   & &\left.
  - e^{2 {\sf i} \pi^2 t \left(2 m^2-4 m n+3 n^2\right)} (-2 R(n,\mu ,\nu )+R(n,\mu,\nu +1)+R(n,\mu +1,\nu ))
   \right)
   \;,
\end{eqnarray}
where the abbreviation
\begin{equation}\label{SA8}
  R(k,\mu,\nu)\equiv \text{erfi}\left(\frac{(1+i) (2 k \pi  t-\mu +\nu )}{2 \sqrt{t}}\right)
\end{equation}
has been used. In the case $m=n$ we have
\begin{eqnarray}\nonumber
   \langle \mu,n|\nu,n,t\rangle&=&\frac{e^{-2 {\sf i} \pi ^2 n^2 t}}{2 \sqrt{\pi }}
   \left(
  (1+{\sf i}) \sqrt{t} \left(e^{\frac{{\sf i} (\mu -\nu -2 \pi  n t+1)^2}{2 t}}
  -2 e^{\frac{{\sf i}(-\mu +\nu +2 \pi  n t)^2}{2 t}}+e^{\frac{{\sf i} (-\mu +\nu +2 \pi  n t+1)^2}{2 t}}\right)\right.\\
  \nonumber
  &&\left.-{\sf i} \sqrt{\pi } \left(-2 (-\mu +\nu +2 \pi  n t) R(n,\mu ,\nu )+(-\mu +\nu
   +2 \pi  n t+1) R(n,\mu ,\nu +1)\right.\right.\\
   \label{SA9}
   &&\left.\left.+(-\mu +\nu +2 \pi  n t-1) R(n,\mu +1,\nu )\right)
   \right)
   \;.
\end{eqnarray}

\begin{figure}[t]
\centering
\includegraphics[width=0.7\linewidth]{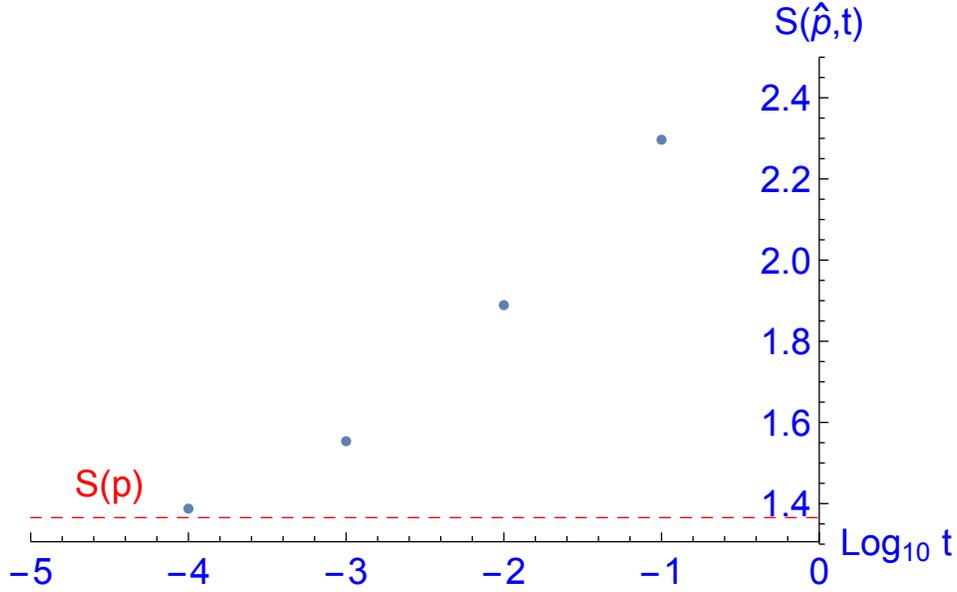}
\caption{Illustration of the increase of the Shannon entropies, $S(p)<S(\hat{p},t)$, in the case of two sequential L\"uders measurements. $t$ denotes the
time difference between the two measurements and assumes the dimensionless values $10^{-4},\ldots, 10^{-1}$;
the parameter $\sigma$ in (\ref{SA3}) is chosen as $\sigma=1$. The entropy after the first measurement is $S(p)=1.3654$ (dashed red line).}
\label{FIG2nd}
\end{figure}

We have noted in subsection \ref{sec:SMM} that in general the $p(j|i)$ need not be symmetric. Indeed, for our example we find, e.~g., that
for $t=\sigma=1$ we have $p(1,1|0,0)=0.00483946$, but  $p(0,0|1,1)=0.00258997$.

The equations (\ref{SA3a}) -- (\ref{SA9}) yield the second marginal probabilities $\hat{p}(j)=\sum_i p(j|i)\,p(i)$
in the form of a doubly infinite series of terms given by analytical expressions. Hence the $\hat{p}(j)$ can be numerically calculated by
a suitable truncation of the infinite series. Moreover, the Shannon entropies $S(p)$ and $S(\hat{p})$ can be numerically calculated
and the $2^{nd}$ law-like statement (\ref{S3}) can be tested, see Figure \ref{FIG2nd}. Due to the localization of the particle by the first measurement
an additional spreading of the momenta is generated that leads to a stronger increase of the Shannon entropy than the increase that is solely produced
by the spreading of the Gaussian wave packet (\ref{SA3}) in the course of time.

%%%%%%%%%%%%%%%%%%%%%%%%%%%%%%%%%%%%%%%%%%%%%%%%%%%%%%%%%%%%%%%%%%%%%%%%%%%%%%%%%%%%%%%%%%%%%%%%%%%%%%%%%%%%%%%%%%%%%%%%%%%%%%%
\subsection{Crooks fluctuation theorems}\label{sec:F}
%%%%%%%%%%%%%%%%%%%%%%%%%%%%%%%%%%%%%%%%%%%%%%%%%%%%%%%%%%%%%%%%%%%%%%%%%%%%%%%%%%%%%%%%%%%%%%%%%%%%%%%%%%%%%%%%%%%%%%%%%%%%%%%

In the literature the Jarzynski equation is sometimes derived from so-called Crooks fluctuation theorems, see \cite{CHT11}.
The notation refers to \cite{C99} where G.~E.~Crooks proved a classical work fluctuation theorem. Quantum versions
of the Crooks fluctuation theorem have first been considered in \cite{K00} and \cite{T00}.
One may ask whether the quantum version of a Crooks fluctuation theorem has a counterpart in the (modified) statistical model of sequential measurements
and whether one would need additional assumptions to prove it.

We adopt the notation of subsection \ref{sec:SMM} and will additionally only assume that the hypothetical probabilities are positive,
$q(j)>0$ for all $j\in{\mathcal J}$. Then we may define a ``reciprocal model of sequential measurements"
$({\mathcal J}, {\mathcal I}, \tilde{P}, D, d)$ as follows.
\begin{equation}\label{F1}
  \tilde{P}(j,i)\equiv \tilde{\pi}(i|j)\,q(j)\equiv \frac{\pi(j|i)\,d(i)}{D(j)}\,q(j)\quad\mbox{for all } j\in{\mathcal J}\mbox{ and } i\in{\mathcal I}
  \;.
\end{equation}
We set $\tilde{\sf E}\equiv {\mathcal J}\times {\mathcal I}$.
First one shows that $\sum_{i}\tilde{P}(j,i)=q(j)$  and $\sum_{i,j}\tilde{P}(j,i)=1$
by means of $\pi$ being a modified doubly stochastic matrix, see (\ref{SM28}). Hence
$q(j)$ is the first marginal probability and $ \tilde{\pi}(i|j)$ the first conditional probability of $\tilde{P}$.
Next one proves that $ \tilde{\pi}(i|j)$ will also be a modified doubly stochastic matrix, with $d$ and $D$ interchanged:
\begin{equation}\label{F2}
 \sum_j  \tilde{\pi}(i|j)\,D(j) \stackrel{(\ref{F1})}{=} \sum_j  \pi(j|i)\,d(i) = d(i) \mbox{ for all } i\in{\mathcal I}
 \;.
\end{equation}
It follows that $({\mathcal J}, {\mathcal I}, \tilde{P}, D, d)$ satisfies a ``reciprocal J-equation" that will be, however, not needed in what follows.

We rather consider the original J-equation (\ref{P2b}) and define the corresponding random variable $Y$ (in a less sloppy way than above) as
\begin{equation}\label{F3}
Y(i,j)=\frac{d(i) q(j)}{D(j) p(i)} \mbox{  for all } i\in{\mathcal I} \mbox{ and } j\in{\mathcal J}
\;.
\end{equation}
We then  rewrite the expectation value of  $Y$ in the following way.
For any real number $y\in{\mathbbm R}$ we define
\begin{equation}\label{F4}
  {\sf E}_y \equiv \{(i,j)\in{\sf E}\left|Y(i,j)=y\right.\}
  \;.
\end{equation}
Let ${\sf Y}\equiv \{y\in{\mathbbm R}\left| {\sf E}_y \neq \emptyset\right.\}$, then
\begin{equation}\label{F5}
  \left\langle Y \right\rangle = \sum_{y\in{\sf Y}}\left(\sum_{(i,j)\in {\sf E}_y} P(i,j) \right)y
  \;.
\end{equation}
The sum in the brackets can be interpreted as the probability that $Y$ assumes the value $y$, or, in symbols:
\begin{equation}\label{F6}
 P(Y=y)=\sum_{(i,j)\in {\sf E}_y} P(i,j)
 \;.
\end{equation}
Next we repeat the above definitions for the reciprocal model $({\mathcal J}, {\mathcal I}, \tilde{P}, D, d)$
setting
\begin{equation}\label{F7}
  \tilde{Y}(j,i)=\frac{D(j)p(i)}{d(i)q(j)}=\frac{1}{Y(i,j)}
  \;,
\end{equation}

\begin{equation}\label{F8}
  \tilde{\sf E}_z=\{(j,i)\in \tilde{\sf E}\left| \tilde{Y}(j,i)=z\right.\}
  \;,
\end{equation}
for $z\in{\mathbbm R}$
and
\begin{equation}\label{F9}
 \tilde{P}(\tilde{Y}=1/y)=\sum_{(j,i)\in \tilde{\sf E}_{1/y}} \tilde{P}(j,i)
 \;.
\end{equation}
We note that
\begin{equation}\label{F10}
 (i,j)\in{\sf E}_y \Leftrightarrow (j,i)\in \tilde{\sf E}_{1/y}
 \;,
\end{equation}
and formulate the following
\begin{prop}\label{Prop4}
For all $y\in{\sf Y}$ there holds
\begin{equation}\label{F11}
  \frac{P(Y=y)}{\tilde{P}(\tilde{Y}=1/y)}=\frac{1}{y}
  \;.
  \end{equation}
\end{prop}
\noindent {\bf Proof}:
\begin{eqnarray}
\label{F12a}
{\tilde{P}(\tilde{Y}=1/y)} &\stackrel{(\ref{F9})}{=}&
 {\sum_{(j,i)\in \tilde{\sf E}_{1/y}} \tilde{P}(j,i)}\\
 \label{F12b}
  &\stackrel{(\ref{F1},\ref{F10})}{=}&  {\sum_{(i,j)\in {\sf E}_y} \pi(j|i)\frac{d(i)q(j)}{D(j)}}\\
 \label{F12c}
  &\stackrel{(\ref{F3},\ref{F4})}{=}&  {\sum_{(i,j)\in {\sf E}_y} \pi(j|i)p(i)\,y}\\
 \label{F12d}
  &\stackrel{(\ref{SM7})}{=}& \left(\sum_{(i,j)\in {\sf E}_y} P(i,j)\right)\,y\\
 \label{F12e}
 &\stackrel{(\ref{F6})}{=}&P(Y=y)\,y
 \;.
\end{eqnarray}
From this the proposition follows immediately. \hfill$\Box$\\

From Proposition \ref{Prop4} we may again derive the J-equation (\ref{P2b}) in the following way:
\begin{equation}\label{F13}
 \left\langle Y \right\rangle
 \stackrel{(\ref{F5},\ref{F6})}{=}
  \sum_{y\in{\sf Y}} P(Y=y) y
 \stackrel{(\ref{F11})}{=}
\sum_{y\in{\sf Y}}\tilde{P}(\tilde{Y}=1/y)=1
 \;.
\end{equation}

It remains to show that (\ref{F11}) indeed entails the Crooks fluctuation theorem in the  case of quantum mechanics.
To this end we assume the case of subsection \ref{sec:L} with $N=1$ such that
\begin{equation}\label{F14}
  Y(i,j)= \exp\left( -\beta \left(W(i,j)-\Delta F\right)\right)
  \;,
\end{equation}
where we have denoted the random variable ``work" by a capital letter $W$,
and hence, writing $y=\exp\left( -\beta \left(w-\Delta F\right)\right)$,
\begin{equation}\label{F15}
 P(Y=y)=P(W=w)
 \;.
\end{equation}
Analogously,
\begin{equation}\label{F16}
 \tilde{P}(\tilde{Y}=1/y)=\tilde{P}(\tilde{W}=-w)
 \;,
\end{equation}
and the Crooks fluctuation theorem assumes its familiar form
\begin{equation}\label{F17}
  \frac{P(W=w)}{\tilde{P}(\tilde{W}=-w)}=\exp\left(\beta \left(w-\Delta F\right)\right)
  \;.
  \end{equation}

Next we will investigate the question how the reciprocal model $({\mathcal J}, {\mathcal I}, \tilde{P}, D, d)$
can be realized in quantum theory. We consider its conditional probability
\begin{equation}\label{F18}
 \tilde{\pi}(i|j)\stackrel{(\ref{F1})}{=}\pi(j|i)\frac{d(i)}{D(j)}\stackrel{(\ref{G21})}{=}
 \mbox{Tr}\left(  \utilde{Q}_j\,U\,\frac{\utilde{P}_i}{d(i)}\,U^\ast  \right)\,\frac{d(i)}{D(j)}
 =
  \mbox{Tr}\left(  \utilde{P}_i\,U^\ast\,\frac{\utilde{Q}_j}{D(j)}\,U  \right)
  \;,
\end{equation}
where the last equation was obtained by cyclic permutation of the operators inside the trace.
This suggests the following realization: We prepare a state described by a statistical operator
$\tilde{\rho}$ such that
\begin{equation}\label{F19}
 q(j)=\mbox{Tr}\left( \tilde{\rho}\,\utilde{Q}_j\right)
 \;,
\end{equation}
and the assumption
\begin{equation}\label{F20}
   \utilde{Q}_j\, \tilde{\rho}\,\utilde{Q}_j=\frac{q(j)}{D(j)}\utilde{Q}_j \mbox{ for all } j\in{\mathcal J}
    \;,
  \end{equation}
analogous to (\ref{AS2a}) is satisfied. Then at the time $t=t_0$ we measure the
set of observables described by the mutually commuting self-adjoint operators $\utilde{F}_1,\ldots,\utilde{F}_L$
with common eigenprojections $\utilde{Q}_j,\;j\in{\mathcal J}$, the measurement being of L\"uders type.
Thereafter a unitary time evolution $\tilde{U}\equiv U^\ast$ takes place until $t=t_1$ where a second measurement
of $\utilde{E}_1,\ldots,\utilde{E}_L$
with common eigenprojections $\utilde{P}_i,\;i\in{\mathcal I}$ is performed. The corresponding conditional probabilities
of the sequential measurements are then given by (\ref{F18}).

In an experiment it might be difficult to realize the adjoint time evolution $\tilde{U}\equiv U^\ast$. As a more practical
alternative we briefly recapitulate the time evolution considered in \cite{CHT11}, section IV A, using our own notation.
Let the time evolution $U(t,t_1)$ for $t\in[t_0,t_1]$ of the original model be given as the solution of the differential
equation
\begin{equation}\label{F21}
  \frac{\partial }{\partial t}U(t,t_1) = -{\sf i}\, H(t)\, U(t,t_1)
  \;,
\end{equation}
with ``initial" value $U(t_1,t_1)={\mathbbm 1}$. Then consider the Hamiltonian
\begin{equation}\label{F22}
\check{H}(t)\equiv H(t_1+t_0-t)
\end{equation}
according to a ``time-reversed protocol" and the corresponding evolution operator $\check{U}(t,t_1)$ satisfying
\begin{equation}\label{F23}
  \frac{\partial }{\partial t}\check{U}(t,t_1) = -{\sf i}\, \check{H}(t)\,\check{U}(t,t_1)
  \;,
\end{equation}
with initial value $\check{U}(t_1,t_1)={\mathbbm 1}$. Upon the transformation $t\mapsto t_1+t_0-t$  Eq.~(\ref{F23}) assumes the form
\begin{equation}\label{F24}
  \frac{\partial }{\partial t}\check{U}(t_1+t_0-t,t_0) ={\sf i}\, \check{H}(t_1+t_0-t)\,\check{U}(t_1+t_0-t,t_0)
  \stackrel{(\ref{F22})}{=}{\sf i}\, H(t)\,\check{U}(t_1+t_0-t,t_0)
  \;.
\end{equation}
Now assume ``micro-reversibility", i.~e., the existence of an anti-unitary operator $\Theta$ commuting with all $H(t)$:
\begin{equation}\label{F25}
 \Theta^\ast\,H(t)\,\Theta = H(t) \mbox{ for all } t\in[t_0,t_1]
 \;,
\end{equation}
and further
\begin{equation}\label{F25a}
 \Theta^\ast\,\utilde{Q}_j\,\Theta = \utilde{Q}_j \mbox{ for all } j\in {\mathcal J}
 \mbox{ and }
 \Theta^\ast\,\utilde{P}_i\,\Theta = \utilde{P}_i \mbox{ for all } i\in {\mathcal I}
 \;.
\end{equation}
The latter assumption already follows from (\ref{F25}) if the $ \utilde{P}_i$ are the eigenprojections of $H(t_0)$ and the
$ \utilde{Q}_j$  the eigenprojections of $H(t_1)$. Our assumption of micro-reversibility is somewhat weaker than the usual
formulation in so far as it only requires that there exist some basis such that $H(t)$ is real for all $t$, see \cite{SKCSG18}
for a similar approach.

Accordingly (\ref{F24}) implies
\begin{equation}\label{F26}
  \frac{\partial }{\partial t}\Theta^\ast\,\check{U}(t_1+t_0-t,t_0)\,\Theta =-{\sf i}\, H(t)\,\Theta^\ast\,\check{U}(t_1+t_0-t,t_0)\,\Theta
  \;.
\end{equation}
Comparison with (\ref{F21}) together with the initial conditions yields
\begin{equation}\label{F27}
  U(t,t_1) = \Theta^\ast\,\check{U}(t_1+t_0-t,t_0)\,\Theta
  \;,
\end{equation}
cp.~Eq.~$(40)$ in \cite{CHT11},
especially
\begin{equation}\label{F28}
U^\ast= U(t_1,t_0)^\ast= U(t_0,t_1) = \Theta^\ast\,\check{U}(t_1,t_0)\,\Theta
  \;.
\end{equation}
Inserting this result into (\ref{F18}) and using (\ref{F25a}) gives
\begin{equation}\label{F29}
 \tilde{\pi}(i|j)
 =
  \mbox{Tr}\left(  \utilde{P}_i\,\check{U}(t_1,t_0)\,\frac{\utilde{Q}_j}{D(j)}\,\check{U}(t_1,t_0)^\ast  \right)
  \;,
\end{equation}
and thus shows that the time-reversed protocol correctly realizes the reciprocal model. But we stress that the
assumptions of micro-reversibility are convenient but not necessary for the validity of the Crooks fluctuation theorem in contrast to the impression
generated by \cite{CHT11}.

As a special case we mention the situation where $H(t)=\check{H}(t)=H(t_1+t_0-t)$
and $\utilde{Q}_i=\utilde{P}_i$ for all $i\in{\mathcal I}={\mathcal J}$, but all preceding assumptions still hold,
in particular (\ref{F25}) and (\ref{F25a}). This includes the case where $H$ is time-independent. Then it follows that
$U^\ast=\Theta^\ast\,U\,\Theta$ and hence
\begin{equation}\label{F30}
 \pi(j|i)\,d(i)=\mbox{Tr}\left(\utilde{P}_j\,U\,\utilde{P}_i\,U^\ast\right)
 =\mbox{Tr}\left(\utilde{P}_i\,\Theta^\ast\,U\,\Theta\,\utilde{P}_j\,\Theta^\ast\,U^\ast\,\Theta\right)=
 \mbox{Tr}\left(\utilde{P}_i\,U\,\utilde{P}_j\,U^\ast\right)= \pi(i|j)\,d(j)
 \;.
\end{equation}
This means that the symmetry condition (\ref{SM31}) considered by W.~Pauli will be exactly satisfied, not only in the Golden Rule approximation.
In the counter-example to (\ref{SM31}) in subsection \ref{sec:SA} the condition (\ref{F25a}) is violated since the momentum
$p_n$ is inverted under time reflections.

%%%%%%%%%%%%%%%%%%%%%%%%%%%%%%%%%%%%%%%%%%%%%%%%%%%%%%%%%%%%%%%%%%%%%%%%%%%%%%%%%%%%%%%%%%%%%%%%%%%%%%%%%%%%%%%%%%%%%%%%%%%%%%%
\section{Applications to classical theory}\label{sec:C}
%%%%%%%%%%%%%%%%%%%%%%%%%%%%%%%%%%%%%%%%%%%%%%%%%%%%%%%%%%%%%%%%%%%%%%%%%%%%%%%%%%%%%%%%%%%%%%%%%%%%%%%%%%%%%%%%%%%%%%%%%%%%%%%

In classical statistical mechanics all observables have definite values for each individual system.
Hence it is not necessary to adopt the scenario of sequential measurements in the context of Jarzynski equations.
Nevertheless, the statistical model of sequential measurements introduced in Section  \ref{sec:SM} can be useful if suitably re-interpreted.
To this end we set ${\mathcal I}={\mathcal J}={\mathcal X}$ where ${\mathcal X}$ is the $2N$-dimensional phase space of the system under consideration.
Summations over ${\mathcal I}$ or ${\mathcal J}$ will be replaced by integrations using the canonical volume form
$dx=dp_1\ldots dp_N\,dq_1\ldots dq_N$ on ${\mathcal X}$. At the time $t=t_0$ the state of the system will be
described by a probability distribution $p:{\mathcal X}\rightarrow{\mathbbm R}_+$ satisfying
\begin{equation}\label{C1}
  p(x)>0 \mbox{ for all } x\in{\mathcal X}
\end{equation}
and
\begin{equation}\label{C2}
 \int_{\mathcal X}p(x)\,dx=1
 \;.
\end{equation}
The time evolution between $t=t_0$ and $t=t_1$ is deterministic and will be described by a volume preserving map
\begin{equation}\label{C3}
  U:{\mathcal X}\rightarrow{\mathcal X}
  \;.
\end{equation}
Let $A:{\mathcal X}\times {\mathcal X}\rightarrow{\mathbbm R}$ be a random variable. Its expectation value
will be defined by
\begin{equation}\label{C4}
\langle A\rangle \equiv \int_{{\mathcal X}\times{\mathcal X}}dx\,dy\,p(x)\, \delta(y,U(x))\,A(x,y)
= \int_{\mathcal X} dx\,p(x)\,A(x,U(x))
\;.
\end{equation}
In the special case of $A(x,y)=\frac{q(y)}{p(x)}$, where $q:{\mathcal X}\rightarrow{\mathbbm R}_+$ is assumed to satisfy
\begin{equation}\label{C5}
 \int_{\mathcal X}q(y)\,dy=1
 \;,
\end{equation}
we conclude
\begin{eqnarray}
\label{C6a}
  \left\langle \frac{q(y)}{p(x)}\right\rangle &\stackrel{(\ref{C4})}{=}& \int_{\mathcal X} dx\,p(x)\, \frac{q(U(x))}{p(x)} \\
  \label{C6b}
   &=& \int_{\mathcal X} dx\,q(U(x)) \\
   \label{C6c}
   &=&\int_{\mathcal X} dy\,q(y)\stackrel{(\ref{C5})}{=}1
   \;,
\end{eqnarray}
using that $U$ is volume preserving in (\ref{C6c}).
Hence the Eq.~(\ref{G24}) has a classical counterpart and the general Jarzynski equation also holds classically,
in particular for the examples treated in the subsections \ref{sec:L} -- \ref{sec:R}.

%%%%%%%%%%%%%%%%%%%%%%%%%%%%%%%%%%%%%%%%%%%%%%%%%%%%%%%%%%%%%%%%%%%%%%%%%%%%%%%%%%%%%%%%%%%%%%%%%%%%%%%%%%%%%%%%%%%%%%%%%%%%%%
\section{Summary and outlook}\label{sec:SO}
%%%%%%%%%%%%%%%%%%%%%%%%%%%%%%%%%%%%%%%%%%%%%%%%%%%%%%%%%%%%%%%%%%%%%%%%%%%%%%%%%%%%%%%%%%%%%%%%%%%%%%%%%%%%%%%%%%%%%%%%%%%%%%%

The usual formulation of the quantum Jarzynski equation applies to closed systems that are initially in thermal equilibrium,
described by the canonical ensemble, and then subject to two sequential energy measurements. Between the two measurements the system
may be arbitrarily disturbed under the influence of a time-dependent Hamiltonian. The present work can be understood as a gradual
generalization of this situation. Some of these generalizations have already be considered in the literature, see
\cite{T00}, \cite{TMYH13} and \cite{SS07} - \cite{YKT12}, but now they appear in a coherent way as results of a unified approach.
First, we allow for equilibrium scenarios that are rather described by micro-canonical or grand canonical ensembles.
In the next step, we also consider local equilibria, i.~e., $N$ subsystems that have initially different temperatures. This case is treated
in the present paper for the case of local canonical ensembles but the extension to the case of local micro-canonical or grand canonical ensembles is straightforward. Moreover, it turns out that for the general quantum Jarzynski equation the restriction to {\it energy} measurements is no longer necessary. The only essential assumptions are those postulating that the first measurement is of L\"uders type and, additionally, that the state resulting after this measurement is diagonal in the
eigenbasis of the measured observables (Assumption $2$). An example, where this more general point of view is crucial, is the sequential measurement
of the ``quasi-energy" in the case of periodic thermodynamics. This example is briefly touched in our paper but could be expanded w.~r.~t.~results
on the dissipated heat obtained in \cite{SSH19}.

At this point a further natural generalization suggests itself, namely the replacement
of the two sequential (projective) measurements by more general ones described by POV measures and involving more general state transformations than those of L\"uders type, see  \cite{MT11}, \cite{GS14} and \cite{CPF17} for related approaches. It turns out that the simple form of the general Jarzynski equation and of the resulting $2^{nd}$ law-like statements will be lost upon this generalization.
In the paper at hand we have followed a different route of generalization by analyzing the probabilistic core of the general Jarzynski equation.
The result is what we have called a ``statistical model of sequential measurements" that does not explicitly presuppose quantum mechanics and includes
the ``J-equation", cf.~Prop.\ref{Prop3}, as a progenitor of the general quantum Jarzynski equation.
Another benefit of the abstract statistical model is to make clear that the J-equation will exactly
hold even if the correct quantum time evolution is replaced by an approximation, e.~g., the Golden Rule approximation, as far as
the modified doubly stochasticity is retained.
The mathematical clue to prove the J-equation is the assumption of a modified doubly stochastic transition probability that is satisfied in quantum theory and breaks the time-reflection symmetry of the model. Consequently, a $2^{nd}$ law-like statement follows that is different from those mentioned above and joins the theory to previous approaches to the Second Law going back to W.~Pauli and G.~D.~Birkhoff. We have illustrated this result by an example involving discrete position-momentum measurements and describing the spreading of an initial Gaussian wave-packet. The arrow of time remains mysterious but the two arrows arising in thermodynamics and quantum measurement theory point into the same direction.

\begin{acknowledgments}
This work has been funded by the Deutsche
Forschungsgemeinschaft (DFG), Grant No. 397107022
(GE 1657/3-1) within the DFG Research Unit FOR 2692
We thank all members of this research unit, especially Andreas Engel, for stimulating and
insightful discussions and hints to relevant literature.
\end{acknowledgments}

%\newpage

\end{document}